\documentclass[pra,twocolumn,a4paper,showpacs]{revtex4}
\usepackage{latexsym}
\usepackage{amsmath}
\usepackage{amsfonts}
\usepackage{graphicx}
\usepackage{exscale}
\usepackage{amssymb}
\usepackage{bbold,mathrsfs}
\usepackage{psfrag}
\usepackage{natbib}
\newcommand{\ket}[1]{\ensuremath{|\,#1\,\rangle}}
\newcommand{\bra}[1]{\ensuremath{\langle\,#1\,|}}

\newcommand{\mf}[1]{\boldsymbol{#1}}

\newcommand{\ve}{\varepsilon}
\newcommand{\eps}{\epsilon}
\newcommand{\vro}{\varrho}
\newcommand{\be}{\begin{equation}}
\newcommand{\ee}{\end{equation}}
\newcommand{\ba}{\begin{eqnarray}}
\newcommand{\ea}{\end{eqnarray}}

\newcommand{\mc}[1]{\ensuremath{\mathcal{#1}}}

\newcommand{\bc}{\begin{center}}
\newcommand{\ec}{\end{center}}
\newcommand{\bi}{\begin{itemize}}
\newcommand{\ei}{\end{itemize}}
\newcommand{\mean}[1]{\ensuremath{ \langle\,#1\, \rangle}}
\newcommand{\means}[1]{\ensuremath{ \langle\,#1\, \rangle_{\text{st}}}}

\newcommand{\meansb}[1]{\ensuremath{ \big\langle\,#1\, \big\rangle_{\text{st}}}}

\begin{document}

\title{Interference in the resonance fluorescence of two incoherently coupled transitions}
\author{Martin Kiffner}
\author{J\"org Evers}
\author{Christoph H. Keitel}
\affiliation{ Max-Planck-Institut f\"ur Kernphysik, 
Saupfercheckweg 1, 69117 Heidelberg, Germany}

\begin{abstract}
The fluorescence light emitted by a 4-level system in $J=1/2$ to $J=1/2$ 
configuration driven by a monochromatic laser field
and in an external magnetic field is studied. 
We show that 
the spectrum of resonance fluorescence emitted on the $\pi$
transitions 
shows a signature of spontaneously generated interference effects. 
The degree of interference in the fluorescence spectrum can be 
controlled by means of the external magnetic field, 
provided that the Land\'e g-factors of the excited and the ground 
state doublet are different.  For a suitably chosen magnetic field strength, 
the  relative weight of the Rayleigh line can be completely suppressed, 
even for low intensities of the coherent driving field. 
The incoherent fluorescence spectrum emitted on the $\pi$ transitions 
exhibits a very narrow peak whose width and weight depends on the 
magnetic field strength. 
We demonstrate that the  spectrum of resonance fluorescence emitted on the $\sigma$ 
transitions  show an indirect signature of interference. 
A measurement  of the relative peak heights  in the spectrum from 
the $\sigma$ transitions allows to determine the branching ratio of the 
spontaneous decay of each excited state into the $\sigma$ channel. 
\end{abstract}

\maketitle

\section{INTRODUCTION}
Since  the emergence of quantum mechanics, quantum interference has been regarded as 
one of the most exciting and intriguing aspects of  quantum theory~\cite{ficek:int}.  
Although interference effects are present in almost all areas of quantum mechanics, some of them 
particularly attracted the attention of many scientists.  
In the following, we will give two examples of  physical systems that are well known in 
the context of interference effects and are both related to the work presented here. 

First of all, we would like to mention   the so-called 
V-system that has been  intensively discussed by theoretical means.  
This  atomic level scheme is comprised of  two near-degenerate 
excited levels  and one ground state, 
and many authors demonstrated that  a rich variety of  
interference effects should be observable in this system. 
These effects include the modification and quenching of spontaneous  
emission~\cite{agarwal:qst,  lee:97,  zhu:95, zhu:96}, and several  
schemes to control spontaneous emission by means of external fields 
have been suggested~\cite{paspalakis:98l, paspalakis:98, keitel:99, plastina:00}. 
Furthermore, it has been shown that  quantum interference 
leads to strong modifications of the spectrum of resonance fluorescence, 
and for suitable parameters the  complete suppression of  
resonance fluorescence is achievable~\cite{cardimona:82, hegerfeldt:92, zhou:96, zhou:97}. 
The emitted fluorescence light also displays  highly non-classical features like extremely strong 
intensity-intensity correlations and squeezing~\cite{swain:00, gao:02}. 

However, all these schemes  rest on the existence of spontaneously generated coherences between the two upper levels 
that can only arise if the  dipole moments between the two upper 
and the lower level are parallel or at least non-orthogonal. 
This requirement is very hard to meet in an experiment, since 
appropriate atomic systems are not known up to now.  
In order to circumvent this problem, an experiment with a 
molecular system has been performed~\cite{xia:96}, but the experimental results 
could not be reproduced yet~\cite{li:00}. 
A recent experiment demonstrates the existence of  spontaneously generated  coherences 
between spin states in quantum dots~\cite{dutt:05}. 

One of the most famous  interference effects  is certainly Young's double-slit experiment,  
especially because it allows  to explore fundamental concepts of quantum mechanics such 
as the principle of complementarity in a very simple setup. Celebrated thought experiments like 
Feynman's light microscope~\cite{feynman:3} and Einstein's recoiling slits~\cite{bohr:2} employ the 
position-momentum uncertainty relation to demonstrate that it is impossible to observe 
the wave and the particle nature of the interfering quantities (for example, electrons or photons) at 
the same time.   In recent years, a proposal by  Scully et. al.~\cite{scully:91} 
gave rise to a lively debate~\cite{storey:94, scully:95,storey:95, wiseman:95, wiseman:97,duerr:98, luis:99} 
on the interrelation between the   principle of complementarity  
and  the position-momentum uncertainty relation. 

\begin{figure}[t!]
\bc
\includegraphics[scale=1]{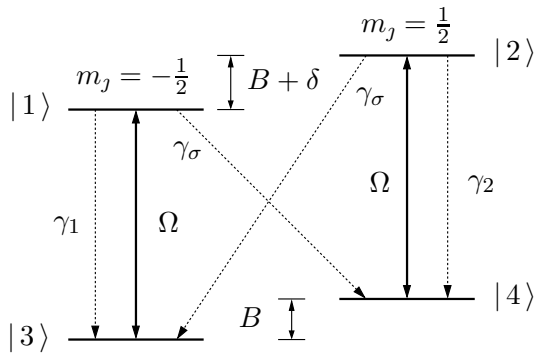}
\caption{\label{picture1} \small Schematic representation of the 
four-level atom of interest. The two upper and  lower levels are Zeeman 
sub-levels  with $m_{\jmath}= \pm\frac{1}{2}$. Each upper state can 
decay by a dipole allowed transition to both ground states. The Zeeman splitting of the magnetic sub-levels 
is not to scale.}
\ec
\end{figure}

A beautiful realization of  Young's two-slit experiment was  performed by Eichmann et. al.~\cite{eichmann:93} 
and subsequently discussed by several authors~\cite{wong:97, itano:98, agarwal:02}.
In this experiment,   the  slits are represented by two ${}^{198}\text{Hg}^+$ ions in a trap 
that are irradiated  by a coherent laser field, and the interference pattern formed by  the scattered light was observed. 
The level scheme of  each of the  two atoms  can be modeled by a $J=1/2$ to $J=1/2$ transition that is also in the 
focus of the work presented here;  
a schematic representation of this  four-level system is shown in Fig.~\ref{picture1}~\cite{polder:76, jakob:99, luetkenhaus:98}. 
The transitions  \mbox{$\ket{1}\leftrightarrow\ket{4}$} and 
\mbox{$\ket{2}\leftrightarrow\ket{3}$} couple 
to $\sigma^+$ and $\sigma^-$ polarized light, respectively, and will be referred  to  as the 
$\sigma$ transitions. By contrast, the $\pi$ transitions  \mbox{$\ket{1}\leftrightarrow\ket{3}$} and 
\mbox{$\ket{2}\leftrightarrow\ket{4}$ } 
couple to light linearly polarized  along $\mf{e}_z$, and their dipole moments are anti-parallel. 
The four-level system of Fig.~\ref{picture1} is  thus  a realistic level scheme with 
non-orthogonal  dipole moments which can be found in real atoms. 
However, it cannot be expected that this four-level system displays the same interference 
effects that were predicted for the V-system with parallel dipole moments since there 
is a striking difference between them.  
In the case of the V-system, both transitions from the upper levels end up in the same ground 
state, while the two $\pi$ transitions of our four-level system start and  end up in different  states 
that are orthogonal to each other.  Thus the question arises whether interference effects can also 
be observed in single $J=1/2$ to $J=1/2$ systems. 

Recently we  investigated  the fluorescence light emitted 
on the $\pi$ transitions in the case of the \textit{degenerate} system (\mbox{$B=\delta=0$} 
in Fig.~\ref{picture1}) and for a monochromatic driving field polarized along $\mf{e}_z$.  
It has been shown~\cite{kiffner:06} that the spectrum of resonance fluorescence  
indeed exhibits a signature of vacuum-mediated interference effects, 
whereas the total intensity is not affected by interference. It has been demonstrated that this 
result is a consequence of the principle of complementarity, applied to time and 
energy.  

Here we generalize our analysis to the non-degenerate system depicted in Fig.~\ref{picture1} 
and include the fluorescence light emitted on the $\sigma$ transitions in our discussion. 
Since it is possible to discriminate between the 
fluorescence light that stems from the $\pi$- and the $\sigma$ transitions 
simply by means of a polarization-dependent 
detection scheme (Sec.~\ref{dscheme}),  these two contributions will be discussed independently 
in Secs.~\ref{sec_pi} and \ref{sigma}, respectively. 

We find that  for the spectrum from the $\pi$ transitions, the degree of 
interference in the coherent and incoherent part of the spectrum strongly depends on the 
frequency difference $\delta$ between the two $\pi$ transitions.  The spectral properties 
of the fluorescence light can thus be controlled by means of an external magnetic field 
that determines the value of $\delta$. 
For example, the relative weight of the Rayleigh line can be completely suppressed for 
a suitable value of $\delta$, even for low intensities of the coherent driving field (Sec.~\ref{coh_sec}). 
The dependence of the incoherent spectrum of resonance fluorescence on the parameter 
$\delta$ is discussed in  Sec.~\ref{inc_sec}. 
Section~\ref{blur} demonstrates  how  the interference terms alter the fluorescence spectrum 
emitted on the $\pi$ transitions for different regimes of the driving field strength. 
The experimental observation of the  spectra including the interference 
terms  could provide evidence for vacuum-mediated interference effects in 
an atomic system. 

The fluorescence spectrum emitted on the $\sigma$ transitions  
of the degenerate system is discussed in Sec.~\ref{sigma}.   
It only consists of an incoherent part and shows an indirect signature of interference  
since the relative peak heights depend on the 
interference terms.  
The measurement  of the relative peak heights 
would also allow to determine the branching ratio of the spontaneous decay of each excited 
state into the  $\sigma$ channel. 

Section~\ref{discussion} provides a detailed discussion of our results. 
In our our previous work~\cite{kiffner:06}, we interpreted 
the interference effect in the spectrum of 
resonance fluorescence in terms of  interferences between transition amplitudes 
that correspond to different time orders of photon emissions. Here we support this 
explanation by a formal argument, and 
the continuous transition from perfect frequency resolution to 
perfect temporal resolution is studied in more detail. 
For a weak driving field and certain values of the parameter $\delta$, the 
incoherent spectra of the $\pi$ and $\sigma$ transitions  
contain a very narrow peak whose width is smaller 
than the natural linewidth. We explain these narrow structures 
in terms of electron shelving. 
Finally, a brief summary of our results is given in Sec.~\ref{sum}.

\section{EQUATION OF MOTION AND DETECTION SCHEME \label{dscheme} }
We now return to the level scheme in Fig.~\ref{picture1}. Note that 
we allow the   Zeeman splitting of the excited and the ground state magnetic sub-levels to be different, 
since the Land\'e g-factors will not necessarily  be the same for these two multiplets. 
For example, in the case of the \mbox{$6s\,{}^2S_{1/2}\,-\,6p\,{}^2 P_{1/2}$} 
transition in ${}^{198}\text{Hg}^+$ the g-factor for the excited states is given by $2/3$, and for the ground states 
it takes on its maximum value of 2. 
The matrix elements of the electric-dipole moment operator $\mf{\hat{d}}$ can be found from the 
the Wigner-Eckart theorem~\cite{sakurai:mqm} and are given by
\ba
 \hspace*{-0.4cm} \mf{d}_1 =\bra{1}\mf{\hat{d}}\ket{3} = - \frac{1}{\sqrt{3}} \,\mc{D}\,\mf{e}_z \, , &&
\mf{d}_2 = \bra{2}\mf{\hat{d}}\ket{4} = -\mf{d}_1 \nonumber \\ 
  \hspace*{-0.4cm} \mf{d}_3 =\bra{2}\mf{\hat{d}}\ket{3} = \sqrt{\frac{2}{3}}\,\mc{D}\,\mf{\eps}^{(-)}\,  , &&
 \mf{d}_4 = \bra{1}\mf{\hat{d}}\ket{4} = \mf{d}_3^* \;. 
\label{dipole}
\ea
In this equation, the circular polarization vector is defined as 
\mbox{$\mf{\eps}^{(-)}=\left(\mf{e}_x - i\,\mf{e}_y\right)/\sqrt{2}$}
and  $\mc{D}$ denotes  the reduced dipole matrix element. 
We assign to  each of the four dipole-allowed transitions a resonance frequency   
$\omega_i$  \mbox{($i \in \{1,2,3,4\}$)}. If the splitting between the 
magnetic sub-levels vanishes (i.e. $B=\delta=0$), these four frequencies are equal.

We are interested in the time evolution of our four level system 
driven by a monochromatic field of frequency $\omega_L$ that is linearly polarized 
along the $z$ axis, 
\be
\mf{E}(t) = E_0\,e^{-i\omega_L t} \,\mf{e}_z \, + \,\text{c.c.} \;, 
\ee
and c.c. stands for the complex conjugate. With this choice of polarization, the electric field couples 
only to the two anti-parallel dipole moments $\mf{d}_1$ and $\mf{d}_2$.  
In the rotating wave approximation, the interaction Hamiltonian 
takes   the form
\be
V =   \big(\,A_{13} -  A_{24}\,\big) \, \hbar \Omega  \,e^{-i \omega_L t} \, + \, \text{h.c.}  \, ,
\ee
where the  atomic transition operators are defined as 
$A_{ij} = \ket{i}\bra{j}$,   and the Rabi frequency is given by 
\mbox{$\Omega =E_0\,\mc{D}/(\sqrt{3}\, \hbar)$}.
The atomic Hamiltonian can  be written as 
\be
H_0 = \hbar\omega_1  \,A_{11} + \hbar(\omega_2+B)\, A_{22} + \hbar B\, A_{44} \;,
\ee 
where $\omega_1$ stands for the resonance frequency of the $1\leftrightarrow 3$ transition and 
$\omega_2=\omega_1+\delta$ is the resonance frequency on the 
$2\leftrightarrow 4$ transition. 
In a rotating frame defined by the unitary transformation 
\be
W = \exp\big[\big(\,A_{11} + A_{22}\,\big)\, i \omega_L t \big]\;,
\label{trafo_w}
\ee
the   master equation for the   density operator $\tilde{\vro} = W \vro W^{\dagger}$  reads 
\be 
\dot{\tilde{\vro}} = - \frac{i}{\hbar}\,[H,\tilde{\vro}]\,+\,\mc{L}_{\gamma}\tilde{\vro}\;. 
\label{master_eq}
\ee
In this equation,  the Hamiltonian is   given by 
\ba
H & = &-\hbar \big[\,\Delta \,A_{11} + (\Delta-\delta)\,A_{22}-B\,(A_{22}+A_{44})\,\big] + \nonumber\\[0.2cm]
& & \big[\,(\,A_{13} -  A_{24}\,)  \, \hbar \Omega  \, + \, \text{h.c.}\,\big] \;, 
\ea
$\Delta = \omega_L - \omega_1$ is the detuning of the driving field from resonance with the 
$1\leftrightarrow 3$ transition, $\Delta-\delta$ is the detuning on the $2\leftrightarrow 4$ transition and the damping term 
$\mc{L}_{\gamma}\tilde{\vro}$ takes the form 
\ba
\mc{L}_{\gamma}\tilde{\vro}  &=&  - \frac{1}{2} \,\sum\limits_{i,\,j=1}^2 \gamma_{ij}\,
\left\{\,[S_i^+,S_j^- \tilde{\vro}] + [\tilde{\vro} S_i^+,S_j^-]\,\right\}\, \nonumber \\
&& - \,\frac{\gamma_{\sigma}}{2} \,\sum\limits_{i=3}^4 
\left\{\,[S_i^+,S_i^- \tilde{\vro}] + [\tilde{\vro} S_i^+,S_i^-]\,\right\}\;.
\label{liou}
\ea
The transition operators $S_i^\pm$    are defined as  
\be
S_1^+ =A_{13}\, , \;\;  S_2^+= A_{24} \, , \; \; S_3^+ = A_{23} \, , \;\;  S_4^+ = A_{14}  \, ,
\label{transitionOp}
\ee
 and $S_i^-=(S_i^+)^{\dagger}$. 
 The decay constant on each of the $\sigma$ transitions 
 is denoted by $\gamma_{\sigma}$, 
the parameters $\gamma_{ij}$ are determined by 
\be
\gamma_{ij}=\frac{\mf{d}_i\cdot\mf{d}_j^*}{|\mf{d}_i|\,|\mf{d}_j|}\sqrt{\gamma_i\gamma_j} 
\qquad i,\,  j \in \{1,2\}\,,
\label{gij}
\ee
and $\gamma_1$ and  $\gamma_2$ are the decay constants of the $\pi$ transitions  (see Fig.~\ref{picture1}). 
For $i=j$, the parameters $\gamma_{ij}$  are equal to the decay rates of the $\pi$ transitions, 
$\gamma_{11}=\gamma_1$ and $\gamma_{22}=\gamma_2$. 
Although  $\gamma_1$ and $\gamma_2$ are  equal in our setup, we will continue to label them differently 
to facilitate the physical interpretation later on. Since $\mf{\hat{d}}_1$ and $\mf{\hat{d}}_2$ 
are anti-parallel, the cross-damping  
terms are given by  \mbox{$\gamma_{12}=\gamma_{21}=-\,\sqrt{\gamma_1\gamma_2}$}. 
These   terms  allow for the possibility of coherence transfer from the excited to the ground state doublet. 

The  decay rates $\gamma_1,\,\gamma_2,\,\gamma_{\sigma}$ can be  related to 
the total decay rate $\gamma=\gamma_1+\gamma_{\sigma}=\gamma_2+\gamma_{\sigma}$ 
of each of the  two excited states through the  branching probabilities $b_{\pi}$ and $b_{\sigma}$, 
\be
\gamma_1=\gamma_2=b_{\pi}\gamma \quad\text{and}\quad \gamma_{\sigma}=b_{\sigma} \gamma\;.
\label{branching_prob}
\ee
According to the Clebsch-Gordan coefficients, we have $b_{\pi}= 1/3$ and  $b_{\sigma}=2/3$. 
Although we will keep the symbols $b_{\pi}$ and $b_{\sigma}$ in formulas, we will always assume these values whenever 
a concrete evaluation is performed, e.g. in figures.

Next we employ the  normalization condition \mbox{$\text{Tr}(\tilde{\vro})=1$} 
to eliminate the matrix element $\tilde{\vro}_{44}$ from 
the master equation  (\ref{master_eq}) that can be cast into the form 
\be
\partial_{t} \mf{R}(t) = \mc{M}\, \mf{R}(t) \, + \, \mf{I}  \,.
\label{bloch_eq}
\ee
Here $\mc{M}$ represents a generalized  $15\times 15$ Bloch matrix, the 
vector  $\mf{I}$ is an inhomogeneity  with components 
\be
\mf{I} = \big(0,0,0,0,0,0,0,i\,\Omega,0,0,0,0,0,-i\,\Omega^*,0\big)^t 
\ee
and the vector $\mf{R}$ contains the matrix elements 
\mbox{$\tilde{\vro}_{ij} =\bra{i}\tilde{\vro}\ket{j}$} 
 of the density operator, 
 \ba
\mf{R} &  = &
\big(\tilde{\vro}_{11} , \tilde{\vro}_{12} , \tilde{\vro}_{13} , 
\tilde{\vro}_{14} , \tilde{\vro}_{21} , \tilde{\vro}_{22} , \tilde{\vro}_{23} , 
\tilde{\vro}_{24} , \label{bloch_vector}  \\[0.2cm]
& & \hspace*{2cm} \tilde{\vro}_{31} , \tilde{\vro}_{32} , 
\tilde{\vro}_{33} , \tilde{\vro}_{34} , \tilde{\vro}_{41} , 
\tilde{\vro}_{42} , \tilde{\vro}_{43}  \big)^t \;. \nonumber
\ea
The stationary solution of Eq.~(\ref{bloch_eq}) is formally   given by
\be
\mf{R}_{\text{st}} = -\mc{M}^{-1}\mf{I} \;, 
\label{psi_stat}
\ee
and an evaluation of the latter equation yields 
\ba
 \tilde{\vro}_{11}& = & 
\frac{1}{2}\,
\frac{|\Omega|^2}{\gamma^2/4 + \delta^2/4+(\Delta-\delta/2)^2 + 2 |\Omega|^2}
 \label{steady}  \\[0.3cm]
\tilde{\vro}_{33} & = & \frac{1}{2}\,
\frac{\gamma^2/4 + \Delta^2+|\Omega|^2}{\gamma^2/4 + \delta^2/4+(\Delta-\delta/2)^2 + 2 |\Omega|^2}  
\nonumber \\[0.3cm]
\tilde{\vro}_{44} & = & \frac{1}{2}\,
\frac{\gamma^2/4 + (\Delta-\delta)^2+|\Omega|^2}{\gamma^2/4 + \delta^2/4+(\Delta-\delta/2)^2 + 2 |\Omega|^2}  
\nonumber \\[0.3cm]
 \tilde{\vro}_{13} & = & \frac{1}{2}\,
 \frac{\left(\Delta - i \gamma/2   \right)\Omega}{\gamma^2/4 + \delta^2/4+(\Delta-\delta/2)^2 + 2 |\Omega|^2}  
 \nonumber \\[0.3cm]
  \tilde{\vro}_{24} & = & \frac{1}{2}\,
 \frac{\left(\delta- \Delta + i \gamma/2   \right)\Omega}{\gamma^2/4 + \delta^2/4+(\Delta-\delta/2)^2 + 2 |\Omega|^2}  \;.
 \nonumber 
 \ea
The remaining non-zero components of $\mf{R}_{\text{st}}$ are  determined by 
 \be
\tilde{\vro}_{11} =  \tilde{\vro}_{22}\; , \qquad \tilde{\vro}_{31}  =   \tilde{\vro}_{13}^* \qquad \text{and}\qquad
\tilde{\vro}_{42}   =   \tilde{\vro}_{24}^* \;.
 \ee
In the case of the degenerate system, the population of the two ground levels will be equal and we have 
\mbox{$\tilde{\vro}_{13}=-\tilde{\vro}_{24}$}.  Note that the minus sign arises since the dipole moments 
$\mf{d}_1$ and $\mf{d}_2$ are anti-parallel, and the  
coherences $\tilde{\vro}_{14}$ and $\tilde{\vro}_{23}$ are equal to zero because the driving field does not 
couple to the $\sigma$ transitions.

In this paper we focus on the total intensity and the spectral distribution of the 
fluorescence light emitted by the atom in steady state.  
The total intensity  
\be
I_{\text{st}}   =   \meansb{\mf{\hat{E}}^{(-)}(\mf{r},t)\cdot\mf{\hat{E}}^{(+)}(\mf{r},t)} 
\label{intensity}
\ee
is  given by the normally ordered first-order correlation function of the 
electric field, and the spectrum of resonance fluorescence 
is determined by the Fourier transform of the two-time correlation function  of 
the electric field~\cite{tannoudji:api},  
\be
S(\omega) = \frac{1}{2\pi} \,\int\limits_{-\infty}^{\infty} e^{-i\omega \tau}\,
\meansb{\mf{\hat{E}}^{(-)}(\mf{r},t+\tau) \cdot\mf{\hat{E}}^{(+)}(\mf{r},t)}\,d\tau \;. 
\label{spectrum}
\ee
In these equations, $\mf{\hat{E}}^{(-)}$  $\left(\mf{\hat{E}}^{(+)}\right)$ denotes the negative 
(positive) frequency part of the electric field operator.  
At a point $\mf{r} = r \mf{\hat{r}}$   in the far-field zone, the negative frequency part 
of the electric field operator  is found to be~\cite{agarwal:qst}
\ba
\mf{\hat{E}}^{(-)}(\mf{r},t)  & = & \mf{\hat{E}}_{\text{free}}^{(-)}(\mf{r},t) \label{field_op} \\[0.2cm]
 & &\hspace*{0.3cm}  -  \frac{\eta}{r}  \, \sum\limits_{i=1}^4\omega_i^2\,
\mf{\hat{r}}\times \left(\,\mf{\hat{r}}\times \mf{d}_i \,\right)\,\tilde{S}_i^+(\hat{t})\,e^{i \omega_L \hat{t}} \;,\nonumber
\ea
where $\hat{t}=t-\frac{r}{c}$ is the retarded time,  $\eta=1/(4\pi\ve_0 c^2)$ 
and \mbox{$\tilde{S}_i^{\pm}=\exp(\mp i\omega_L t)\,S_i^{\pm}$}.  
The first term  stands for the negative frequency part of the 
free field. It does not contribute to 
the normally ordered correlation functions in Eqs.~(\ref{intensity}) and (\ref{spectrum}) 
as long as the point of observation lies outside   
the driving field~\cite{mollow:75}.
The second term describes the retarded dipole field generated by the atom 
situated at the point of  origin.

Throughout this paper we assume that the point of observation lies in 
the $y$-direction, where the $z$- and $x$-axes are defined by the polarization 
and the direction of propagation of the laser beam, respectively. 
An evaluation of the cross products in Eq. (\ref{field_op})  
shows then that the light emitted on the $\pi$ transitions 
is linearly polarized along $\mf{e}_z$, whereas the light emitted on the $\sigma$ transitions 
is linearly polarized along $\mf{e}_x$. The advantage of this detection scheme is that 
one can easily discriminate between the light emitted on the $\pi$ and $\sigma$ transitions 
by means of a polarization filter. For this reason we will discuss the fluorescence light 
of the $\pi$-and $\sigma$ transitions separately. 

\section{SPECTRUM OF RESONANCE FLUORESCENCE -- $\pi$ TRANSITIONS \label{sec_pi}}
\begin{figure}[t!]
\bc
\includegraphics[scale=1]{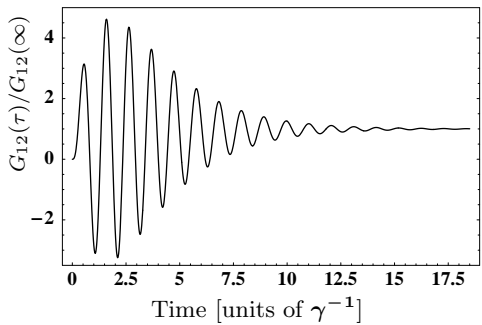}
\caption{\label{picture2} \small Plot of the correlation function $G_{12}$ in 
relation to its long-time limit $G_{12}(\infty)=-\sqrt{\gamma_1\gamma_2}
\means{ \tilde{S}_1^+}\means{  \tilde{S}_2^-}$ for the  degenerate system. The 
 parameters are $\Omega =3\times10^7\,s^{-1}$, \mbox{$\Delta = 5\times 10^6\,s^{-1}$ } 
and \mbox{$\gamma=10^7\,s^{-1}$}. $G_{12}$ has to vanish at $\tau=0$ since the ground states are 
orthogonal.  }
\ec
\end{figure}

We begin with a brief discussion of the 
steady-state intensity recorded by a broadband detector that observes the light emitted  
on the $\pi$ transitions.  
According to Eqs.~(\ref{intensity}) and (\ref{field_op}), we have  
\be
I_{\text{st}}^{\pi}    =  
\phi_{\pi} \,\sum\limits_{i,\,j=1}^{2} \gamma_{ij} \,\means{\tilde{S}_i^+ \tilde{S}_j^- }\,, \label{intensity2}
\ee
where  it was assumed that $\omega_1\approx\omega_2$ to obtain a common prefactor $\phi_{\pi}$ 
that we set equal to one in the following.   
The terms $\gamma_{ij}$ are defined in Eq.~(\ref{gij}), and 
\mbox{$\gamma_{12}=\gamma_{21}=-\,\sqrt{\gamma_1\gamma_2}$} 
describe  the cross-damping  between the $\pi$ transitions that arises 
as a consequence of quantum interference. 
However, these  interference terms do not contribute to the total intensity, 
regardless of what the steady state solution might be, 
because  the ground states are orthogonal,
\be 
\means{\tilde{S}_1^+ \tilde{S}_2^- } = \meansb{\ket{1}\bra{3}\:\ket{4}\bra{2} } = 0 \;.
\label{orthogonal}
\ee
Consequently, the intensity emitted on the 
$\pi$ transitions is not altered by  interference terms  
and simply proportional to the population of the excited states, 
\be
I_{\text{st}}^{\pi}=b_{\pi} \gamma (\tilde{\vro}_{11} + \tilde{\vro}_{22})\;.
\label{intensity4}
\ee
We now turn to the the spectrum of resonance fluorescence emitted 
on the $\pi$ transitions. With the help of Eqs.~(\ref{spectrum}) and  (\ref{field_op})  
we arrive at 
\be
S^{\pi}(\tilde{\omega}) =\frac{1}{\pi} \, \sum\limits_{i,\,j=1}^{2} \gamma_{ij}\, 
\text{Re}\int\limits_{0}^{\infty} e^{-i\tilde{\omega}\tau} 
\means{\tilde{S}_i^+(\hat{t}+\tau)\tilde{S}_j^-(\hat{t})}\, d\tau \;,
\label{specpi}
\ee
where  $\tilde{\omega} = \omega - \omega_L$ is the difference 
between the observed frequency  and the laser frequency. 
In contrast to Eq.~(\ref{orthogonal}), the terms proportional to $\gamma_{12}$  are now 
determined by the \emph{two-time averages} \mbox{$\means{\tilde{S}_1^+(\hat{t}+\tau)\tilde{S}_2^-(\hat{t})}$}  
rather than by the one-time averages.  
Indeed, we find that  the correlation function 
\be
G_{12}(\tau)=-\sqrt{\gamma_1\gamma_2}\,\means{\tilde{S}_1^+(\hat{t}+\tau)\tilde{S}_2^-(\hat{t})}
\label{g12}
\ee
is  different from zero for $\tau > 0$; a plot of  
$G_{12}$  is shown in Fig.~\ref{picture2}.  
But this implies that there is quantum interference in the spectrum of the light 
emitted on the $\pi$ transitions, although there is no interference in the 
total intensity. 
To illustrate this result we 
decompose the transition operators in  Eq.~(\ref{g12}) 
in mean values and fluctuations according to
\be
\tilde{S}_i^{\pm}  =  \means{\tilde{S}_i^{\pm}}\hat{1} \,+\, \delta \tilde{S}_i^{\pm} \;. 
\label{div_op} 
\ee
The correlation function $G_{12}(\tau)$ becomes then 
\ba
G_{12}(\tau) & =& -\sqrt{\gamma_1\gamma_2}\,
\big[\,\means{\delta \tilde{S}_1^+(\hat{t}+\tau)\delta \tilde{S}_2^-(\hat{t})} \\[0.2cm]
& & \hspace*{3cm} + \means{ \tilde{S}_1^+}\means{  \tilde{S}_2^-}\,\big]\,.\nonumber
\ea
The two-time average   of the fluctuations  
can be calculated from the generalized  Bloch equations and  the quantum regression theorem 
(see Appendix). It decays exponentially with a  time constant on the order of  $\gamma^{-1}$ 
and does not contribute to $G_{12}$ in  the long-time limit \mbox{$\tau\rightarrow \infty$}.  
The mean values $\means{\tilde{S}_1^{+}}= \tilde{\vro}_{31}$ and 
$\means{\tilde{S}_2^{+}}= \tilde{\vro}_{42}$ are  
given by matrix elements of the  steady-state density-operator in Eq.~(\ref{steady})
and are both different from zero.  
This is obvious from a physical point of view since the laser field creates a 
coherence on both 
transitions $1\leftrightarrow3$ and $2\leftrightarrow4$. Consequently, 
the long-time limit of $G_{12}$ 
reads \mbox{$G_{12}(\infty) = -\sqrt{\gamma_1\gamma_2}
\means{ \tilde{S}_1^+}\means{  \tilde{S}_2^-}$}.  It follows that the interference terms will affect  
the coherent and incoherent spectrum of resonance fluorescence.  

Before we give expressions for the spectral distribution of the emitted light, we calculate the  
respective contributions  of coherent and incoherent scattering to the intensity $I_{\text{st}}^{\pi}$.  
To this end we apply the decomposition of the transition operators Eq.~(\ref{div_op}) to    Eq.~(\ref{intensity2}).  
This allows us to write    $I_{\text{st}}^{\pi}$   as the sum of four  terms,  
 \be
 I_{\text{st}}^{\pi}=I_{\text{coh}}^0 + I_{\text{coh}}^{\text{int}}+ I_{\text{inc}}^0  + I_{\text{inc}}^{\text{int}}\;.
 \label{intensity3}
\ee
The first two terms account for the contribution of coherent scattering (subscript ``coh'') and are given by 
 \ba
I_{\text{coh}}^0 & = & \gamma_1 \,|\means{\tilde{S}_1^+}|^2 + \gamma_2 \,|\means{\tilde{S}_2^+}|^2  \\[0.2cm]
I_{\text{coh}}^{\text{int}} &=& -2\,  \sqrt{\gamma_1\gamma_2}\;\text{Re}\means{\tilde{S}_1^+ } \means{ \tilde{S}_2^- }    \;.
\label{coherent}
\ea
In this equation,   $I_{\text{coh}}^0$ stands for the contribution of terms proportional to $\gamma_{11}$ and $\gamma_{22}$, and 
 $I_{\text{coh}}^{\text{int}}$ is the weight of the interference terms that can  be positive or negative.    
By contrast, the sum of $I_{\text{coh}}^0$   and $I_{\text{coh}}^{\text{int}}$  is the 
weight of the Rayleigh line that is always positive. 
The last two terms in Eq.~(\ref{intensity3}) denote the contribution of incoherent scattering (subscript ``inc''), 
\ba
I_{\text{inc}}^0 & = & \gamma_1 \,\means{\delta \tilde{S}_1^+ \delta \tilde{S}_1^- } + \gamma_2 
\,\means{\delta \tilde{S}_2^+ \delta \tilde{S}_2^- }  \\[0.2cm]
I_{\text{inc}}^{\text{int}} & =& -2\,  \sqrt{\gamma_1\gamma_2}\;\text{Re}\,
\means{\delta \tilde{S}_1^+ \delta \tilde{S}_2^- } \;. \label{weight}
\ea
Since  the ground states are  orthogonal, Eq.~(\ref{div_op}) allows to  establish the relation 
\be
\means{\delta \tilde{S}_1^+ \delta \tilde{S}_2^- }=- \means{\tilde{S}_1^+ } \means{   \tilde{S}_2^- }   \;.
\ee
If this expression is applied to  Eq.~(\ref{weight}), it follows from Eq.~(\ref{coherent}) that 
the interference terms  $I_{\text{coh}}^{\text{int}}$ and $I_{\text{inc}}^{\text{int}}$ are of 
opposite sign, i.e. 
\be
I_{\text{coh}}^{\text{int}}=- I_{\text{inc}}^{\text{int}} \;.\label{relation}
\ee
This relation clarifies that the   interference terms alter the weights of the coherent and the  incoherent part of the  
spectrum, whereas the total intensity remains unchanged.  
Note that this is in contrast to the  V-system with non-orthogonal transition dipole 
moments mentioned in the introduction, where both the fluorescence spectrum and 
the total intensity show a signature of interference~\cite{ficek:int, lee:97, zhou:96, zhou:97}.

We now turn to the spectral distribution of the fluorescence light and 
employ Eq.~(\ref{div_op}) to write  the spectrum of resonance fluorescence in 
Eq.~(\ref{specpi}) as the sum of  the coherent and the incoherent spectrum, 
\mbox{$S^{\pi} (\tilde{\omega})=S_{\text{coh}}^{\pi} (\tilde{\omega}) +S_{\text{inc}}^{\pi} (\tilde{\omega})$}, 
where 
\ba
\hspace*{-1cm}  S_{\text{coh}}^{\pi} (\tilde{\omega}) & = & \big(I_{\text{coh}}^0 + I_{\text{coh}}^{\text{int}}\big)
\,\delta(\tilde{\omega})  \label{cohpeak} \\[0.2cm]
\hspace*{-1cm} S_{\text{inc}}^{\pi}(\tilde{\omega})  & = & \nonumber \\
& & \hspace*{-1.5cm} \frac{1}{\pi} 
\sum\limits_{i,\,j=1}^{2}  \gamma_{ij} \,
\text{Re}\int\limits_{0}^{\infty} e^{-i\tilde{\omega}\tau}
\means{\delta \tilde{S}_i^+(\hat{t}+\tau) \delta \tilde{S}_j^-(\hat{t}) }\, d\tau \,.
  \label{s_pi}
\ea
These two contributions will be discussed in the following Sections. 

\subsection{Coherent spectrum of resonance fluorescence \label{coh_sec}}
\begin{figure}[t!]
\bc
\includegraphics[scale=1]{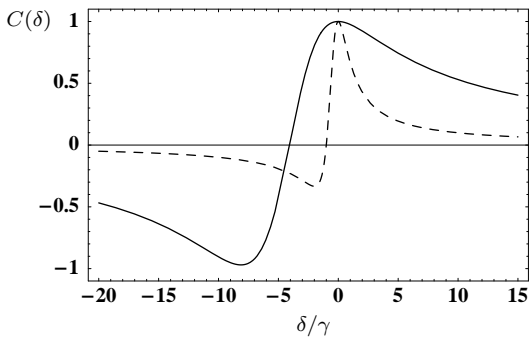}
\caption{\label{picture3} \small   Plot of  the  relative weight of the interference terms 
$C(\delta)$ for different values of the detuning $\Delta$ of the laser field from the $1\leftrightarrow3$ transition. 
The parameters are given by  \mbox{$\gamma=10^7 s^{-1}$},  
\mbox{$\Delta=-4\times10^7 s^{-1}$} (solid line)   
and  \mbox{$\Delta = -5\times 10^6 s^{-1}$} (dashed line). 
}
\ec
\end{figure}

The coherent part of the fluorescence spectrum 
consists of   the Rayleigh peak centered at \mbox{$\omega=\omega_L$}. 
In order to get a better understanding of how the weight of this line is affected by interference, 
we write it as 
\be
I_{\text{coh}}^0 + I_{\text{coh}}^{\text{int}}    =  |\sqrt{\gamma_1}\means{\tilde{S}_1^+}  - 
 \sqrt{\gamma_2}\means{\tilde{S}_2^+}|^2 \;.
\label{w2}
\ee
In this equation,   $\means{\tilde{S}_1^+}$ is proportional to the scattering amplitude 
on the $1\leftrightarrow3$ transition and $-\means{\tilde{S}_2^+}$ corresponds to the scattering amplitude  on the  
$2\leftrightarrow4$ transition.  Note that the minus sign arises since the dipoles $\mf{d}_1$ and $\mf{d}_2$ are 
anti-parallel. Depending on the relative phase and the absolute values of  the coherences $\means{\tilde{S}_1^+}$ 
and $\means{\tilde{S}_2^+}$, there will be constructive or destructive interference in the coherent part of the spectrum.  
We will now demonstrate that  the degree of interference in the coherent spectrum can be controlled by means 
of the difference $\delta$ between the resonance frequencies of the $\pi$ transitions. 
Therefore, we write Eq.~(\ref{w2}) as 
\be
I_{\text{coh}}^0 + I_{\text{coh}}^{\text{int}} = I_{\text{coh}}^0 \,
 \big[\,1+ C \,\big]\;, \label{cohline}
 \ee
where 
\mbox{$C =I_{\text{coh}}^{\text{int}}/I_{\text{coh}}^0$} is the  relative weight of the interference terms. 
An explicit expression for $C$ can be found with  the help of the  definitions in Eq.~(\ref{coherent}) 
and the steady-state solution for $\tilde{\vro}$ in Eq.~(\ref{steady}), 
\be
C =   \frac{\gamma^2/4+\Delta( \Delta-\delta) }{\gamma^2/4 + \delta^2/4+(\Delta-\delta/2)^2 }  \;.
\label{Cfunc}
\ee
The absolute value of this quantity can  be regarded as the degree of interference  in the coherent spectrum. 
Figure~\ref{picture3}  shows  a 
plot of  $C$ as a function of $\delta$ for two different (negative) detunings $\Delta$. 
It is evident  that $C$ is equal to one in the case of the degenerate system.  
Therefore, we have perfect constructive interference  for  
$\delta=0$.  
In this case, the detunings on both $\pi$ transitions are equal and hence 
we have $\means{\tilde{S}_1^+}=-\means{\tilde{S}_2^+}$, the two transitions are now perfectly equivalent.  
In addition, the weight of the Rayleigh line   is then, apart from the 
branching probability $b_{\pi}$,  identical to the corresponding 
expression for a two-level atom~\cite{kimble:76}. 

As $|\delta|$ increases,  $C(\delta)$ decreases monotonously and  becomes zero at 
\mbox{$\delta_0=\Delta\big[1+\gamma^2/(4\Delta^2)\big]$}. Note that $\delta_0$ can be either positive 
or negative, depending on the sign of $\Delta$. 
In the case of   $\Delta^2\gg\gamma^2$, we have $\delta_0\approx\Delta$.  This implies 
that  the interference term vanishes if the laser field 
is resonant with the $2\leftrightarrow4$ transition. 
The minimum of the curve is reached at 
\mbox{$\delta_{\text{min}}=2\Delta(1+\gamma^2/(4\Delta^2))$} 
and given by \mbox{$C(\delta_{\text{min}})= -1/(1+\gamma^2/(2\Delta^2))$}. Consequently,  
$C(\delta_{\text{min}})$ tends to $-1$ provided that $\Delta^2\gg\gamma^2$. 
The weight of the Rayleigh peak becomes  then  zero as a 
consequence of destructive interference, and the emitted radiation is solely incoherent.  
Note that this situation occurs if 
the detunings on the $1\leftrightarrow 3$ and $2 \leftrightarrow 4$ transitions 
are approximately equal and of opposite sign. In this case,  the coherences 
$\means{\tilde{S}_1^+}$ and $\means{\tilde{S}_2^+}$ cancel each other in Eq.~(\ref{w2}). 
Finally, $C$ tends to zero as $|\delta|$ becomes much larger than $|\Delta|$ and $\gamma$. 
This is due to the fact  that the interference term in Eq.~(\ref{coherent}) 
consists of the product of $\means{\tilde{S}_1^+}$ and $\means{\tilde{S}_2^+}$. 
If the detuning on one of the two $\pi$ transitions becomes  very large, $I_{\text{coh}}^{\text{int}}$ 
tends to zero, whereas $I_{\text{coh}}^0$ remains different from zero.

\subsection{Incoherent spectrum of resonance fluorescence \label{inc_sec}}
It is  possible to evaluate the expression for $S_{\text{inc}}^{\pi}$ in 
Eq.~(\ref{s_pi}) analytically, an outline of the calculation  can be found 
in the Appendix. However, the general result is too bulky to present it here. 
We just mention that the spectrum does only depend on the difference  $\delta$ between the 
Zeeman splittings of the ground and excited states, but not on the parameter $B$  
(see Fig.~\ref{picture1}). In the case of the degenerate system, we find 
\be
S_{\text{inc}}^{\pi}(\tilde{\omega}) 
=  b_{\pi}\,\frac{\gamma}{ \pi}\,
\frac{  \gamma^2 + 2 |\Omega|^2 + \tilde{\omega}^2}{\gamma^2/4 +  \Delta^2 + 2|\Omega|^2}
\,\frac{2 \gamma  |\Omega|^4 }{|P(-i \tilde{\omega})|^2} \;,
\label{two_level_spec}
\ee
where $P(z)$ is a cubic polynomial as a function of $z$ that is defined as 
\be
P(z) = \frac{1}{4} (z + \gamma)\big[ (2 z + \gamma)^2 + 4 \Delta^2 \big] + 
2 ( 2 z +\gamma ) |\Omega|^2 \;. 
\ee
Apart from the branching probability $b_{\pi}$,   this result is \textit{exactly} 
the same as the incoherent spectrum of resonance fluorescence of  
a two-level atom~\cite{kimble:76}. 
\begin{figure}[b!]
\bc
\includegraphics[scale=1]{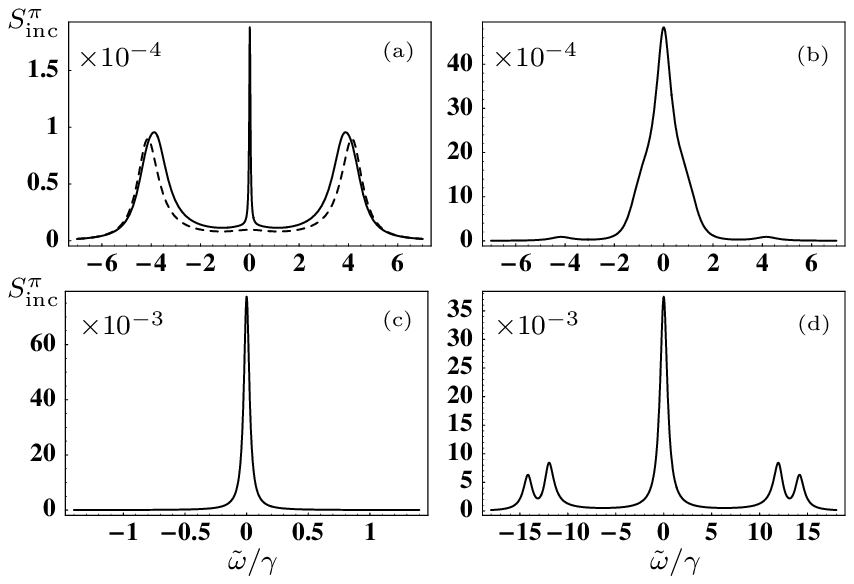}
\caption{ \small  \label{picture4} 
Incoherent spectrum of resonance fluorescence according to Eq.~(\ref{s_pi}). 
Plot  (a)  shows $S_{\text{inc}}^{\pi}$ for the degenerate system (dashed line) and 
for  \mbox{$\delta=-4\times10^6s^{-1}$} (solid line),  
the other parameters are  \mbox{$\gamma=10^7 s^{-1}$},  
\mbox{$\Delta=-4\times10^7 s^{-1}$} and  \mbox{$\Omega=6\times10^6 s^{-1}$}. 
In (b) and (c) the values of $\delta$ are given by \mbox{$\delta=\delta_0$}
and \mbox{$\delta=\delta_{\text{min}}$}, respectively, 
the other parameters are the same than in (a).  
Plot (d) shows the incoherent spectrum for the set of parameters 
\mbox{$\Delta =-5\times10^6 s^{-1}$}, \mbox{$\Omega=6\times10^7 s^{-1}$},
\mbox{$\gamma=10^7 s^{-1}$} and \mbox{$\delta=-8\times10^7 s^{-1}$}. 
}
\ec
\end{figure}

As soon as $\delta$ becomes different from zero, the incoherent spectrum differs 
considerably  from the two-level spectrum. This is demonstrated in 
Fig.~\ref{picture4}~a) which displays  $S_{\text{inc}}^{\pi}$ for $\delta=0$ (dashed line) 
and \mbox{$\delta=-4\times10^6 s^{-1}$} (solid line). For $\delta\not=0$, an additional 
central peak occurs whose width is much smaller than the decay rate $\gamma$.  

Section~\ref{coh_sec} provides a detailed discussion of  the weight of the  
interference term  $I_{\text{coh}}^{\text{int}}$ in the coherent spectrum. 
These results can also be applied to the weight of the  interference term  $I_{\text{inc}}^{\text{int}}$ in the inelastic spectrum 
by means of Eq.~(\ref{relation}). 
For example, it follows  that the weight of the  
interference term  $I_{\text{inc}}^{\text{int}}$ in the inelastic spectrum vanishes  for  $\delta=\delta_0$.  
This situation is shown in Figure ~\ref{picture4}~b),   where 
the width and the weight of the  additional peak is  larger than in a). 
For $\delta=\delta_{\text{min}}$ and the parameters of Fig.~\ref{picture4}, we know from Sec.~\ref{coh_sec} 
that the weight of the Rayleigh line is approximately zero. 
The corresponding incoherent spectrum is shown in Fig.~c). Instead of the elastic delta-peak in the coherent 
spectrum we thus have a very narrow peak that occurs  in the incoherent spectrum. 
\begin{figure}[t!]
\bc
\includegraphics[scale=1]{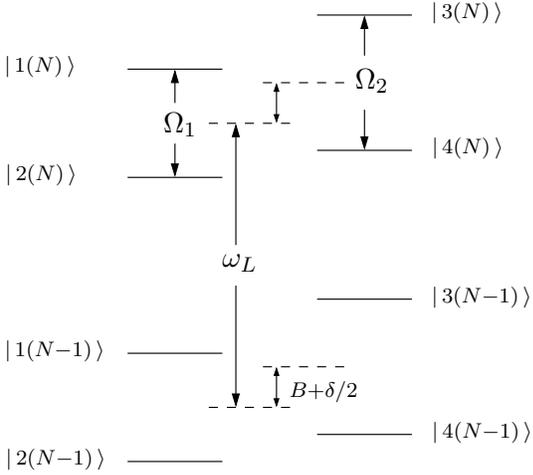}
\caption{\label{picture5} \small Dressed state analog  of the bare state system in Fig.~\ref{picture1}. The 
frequency of the laser field is labeled by $\omega_L$.  For $\delta\not=0$, the detuning of the laser field will be 
different on each of the $\pi$ transitions. There are thus two effective Rabi frequencies $\Omega_1$ and $\Omega_2$ 
involved. The splitting of the dressed states for fixed $N$ is not to scale. }
\ec
\end{figure}

Finally, Fig.~\ref{picture4}~d) shows $S_{\text{inc}}^{\pi}$ for a strong laser field. In this case, 
the weight of the interference terms is  negligible as can be verified with the help of 
Eq.~(\ref{coherent}). However, the incoherent spectrum still deviates from the Mollow spectrum if 
$\delta\not=0$. This can be easily understood with the aid of the dressed states~\cite{tannoudji:api,tannoudji:77} 
of the system. If $N$ denotes the number of laser photons of frequency $\omega_L$,  
the dressed states can be expressed in terms of the bare states as follows,
\ba
\ket{1(N)} & = & e^{i \phi} \sin\Theta_1\,\ket{1,\,N} + \cos \Theta_1\,\ket{3,\,N+1} \quad \nonumber \\[0.1cm]
\ket{2(N)} & = & e^{i \phi} \cos\Theta_1\,\ket{1,\,N} - \sin \Theta_1\,\ket{3,\,N+1} 
\label{dressed1}
\ea
where \mbox{$\tan 2\Theta_1 =  2\,|\Omega|/\Delta$} and 
\ba
\ket{3(N)} & = & e^{i \phi}\sin\Theta_2\,\ket{2,\,N} - \cos \Theta_2\,\ket{4,\,N+1}  \quad \nonumber\\[0.1cm]
\ket{4(N)} & = & e^{i \phi}\cos\Theta_2\,\ket{2,\,N} + \sin \Theta_2\,\ket{4,\,N+1} 
\label{dressed2}
\ea
with \mbox{$\tan 2\Theta_2 = 2\,|\Omega|/(\Delta -\delta)$} ($0 < \Theta_1,\,\Theta_2 < \pi/2$, 
$e^{i \phi}=\Omega/|\Omega|$).  
Figure~\ref{picture5} shows the relative position of the dressed states for two manifolds with $N$ and 
$(N-1)$ laser photons, respectively. Note 
that $\ket{1(N)}$ and $\ket{2(N)}$ are separated by a frequency interval of 
\mbox{$\Omega_1 = \sqrt{4|\Omega|^2 + \Delta^2}$}, whereas 
the spacing between $\ket{3(N)}$ and $\ket{4(N)}$ is given by   
\mbox{$\Omega_2  =\sqrt{4|\Omega|^2 + (\Delta-\delta)^2}$}.  
The sidebands in the spectrum of the $\pi$ transitions result from   the 
transitions \mbox{$\ket{1(N)}\rightarrow\ket{2(N-1)}$}, \mbox{$\ket{2(N)}\rightarrow\ket{1(N-1)}$}, 
\mbox{$\ket{3(N)}\rightarrow\ket{4(N-1)}$} and \mbox{$\ket{4(N)}\rightarrow\ket{3(N-1)}$}. 
Consequently, they will be located at the frequencies 
\mbox{$\omega_L \pm \Omega_1$} and \mbox{$\omega_L \pm \Omega_2$}. For $\delta\not=0$, 
we thus expect four sideband peaks symmetrically placed around the laser frequency $\omega_L$, 
precisely as  in Fig.~\ref{picture4}~d).

\subsection{Influence of the interference terms on the fluorescence spectrum \label{blur}}
In this Section we investigate how the interference terms alter the fluorescence spectrum 
emitted on the $\pi$ transitions.  
Here we only  consider  the degenerate system that is distinguished by maximal constructive 
(destructive) interference in the coherent (incoherent) part of the fluorescence spectrum, 
see Sec.~\ref{sec_pi}. 
If the interference terms  in Eq.~({\ref{specpi}) are omitted, the 
fluorescence spectrum   reads   
\be
S_0^{\pi}(\tilde{\omega}) =\frac{1}{\pi}  \sum\limits_{i=1}^{2} \gamma_{ii} \,\text{Re}
\int\limits_{0}^{\infty} e^{-i\tilde{\omega}\tau} 
\means{\tilde{S}_i^+(\hat{t}+\tau)\tilde{S}_i^-(\hat{t})}\, d\tau \;. \label{specpiperp}  
\ee
The fluorescence spectra with and without the 
interference terms according to Eqs.~(\ref{specpi}) and (\ref{specpiperp}) 
are  shown in Fig.~\ref{picture6}  for   
different parameters of the laser field.  
If the saturation parameter defined in Eq.~(\ref{saturation}) 
is much larger than unity, the weight of the interference terms goes to zero. 
However, Fig.~\ref{picture6}~a) demonstrates that the interference terms 
still alter the shape of the fluorescence spectrum in the region of the sideband peaks. 
The spectrum $S^{\pi}$ with interference terms is identical to the fluorescence 
spectrum of a two-level atom (see Sec~\ref{sec_pi}), and thus the ratio between the central and the 
sideband peaks reads $1:3:1$.  For the spectrum without the interference terms and 
a branching probability of $b_{\pi}=1/3$, this ratio reads~$7:15:7$.  

Figure~\ref{picture6}~b) shows $S^{\pi}$ and $S_0^{\pi}$  for low saturation. 
In this case, the spectrum without interference terms 
is distinguished by a narrow peak centered at the laser frequency that occurs in 
addition to the elastic Rayleigh peak. A numerical analysis shows that 
$S_0^{\pi}$ can be written as 
\be 
S_0^{\pi}(\tilde{\omega})  \approx   I_{\text{coh}}^0 
\,\delta(\tilde{\omega}) \,
   +\, S_{\text{inc}}^{\pi}(\tilde{\omega}) \,+ \,S_{\text{peak}}^{\pi}(\tilde{\omega})\,.  
   \label{specpiperpapprox}
\ee
In this equation, the first term represents the Rayleigh peak whose weight misses the  
interference term  $I_{\text{coh}}^{\text{int}}$ that is  present in 
Eq.~(\ref{cohpeak}).  The  second term stands for 
the  incoherent spectrum according to Eq.~(\ref{two_level_spec}). 
The last term describes a Lorentzian of weight  
$I_{\text{coh}}^{\text{int}}$ and 
width $\Gamma_{\pi}$ that is centered at the laser frequency, 
\be
S_{\text{peak}}^{\pi}(\tilde{\omega}) = \frac{I_{\text{coh}}^{\text{int}}}{\pi}\,
\frac{ \Gamma_{\pi} }{\tilde{\omega}^2 + \Gamma_{\pi}^2}\;. \label{peak}
\ee
The weight of the extra peak $S_{\text{peak}}^{\pi}$ is 
determined by the constraint that the total intensity is independent of 
the interference terms (see Sec.~\ref{sec_pi}). 
Therefore,   $S_{\text{peak}}^{\pi}$  has to 
compensate for the reduced weight of the Rayleigh line of $S_0^{\pi}$ as compared 
to the spectrum with interference terms. 
In general, the width $\Gamma_{\pi}$ of the extra peak $S_{\text{peak}}^{\pi}$ is  smaller than the decay 
rate $\gamma$.  If the  saturation parameter $s$   
is much smaller than unity, we find  ($b_{\pi}=1/3$) 
\be
I_{\text{coh}}^{\text{int}}  \approx  \frac{\gamma}{12} (1-2 s)s \quad\text{and}\quad
\Gamma_{\pi}  \approx  2\frac{\gamma}{9} (3 - 5 s)s \,.
\label{gamma_pi}
\ee
Figure~\ref{picture6}~b) allows to summarize the effect of the interference terms on the 
fluorescence spectrum in the case of low saturation as follows. The spectrum 
without interference terms displays a  narrow peak $S_{\text{peak}}^{\pi}$ 
of finite width at the laser frequency that is absent if the interference terms are 
taken into account. Therefore, quantum interference cancels the 
incoherent response of the atom at the laser frequency $\omega_L$.    
\begin{figure}[t!]
\bc
\includegraphics[scale=1]{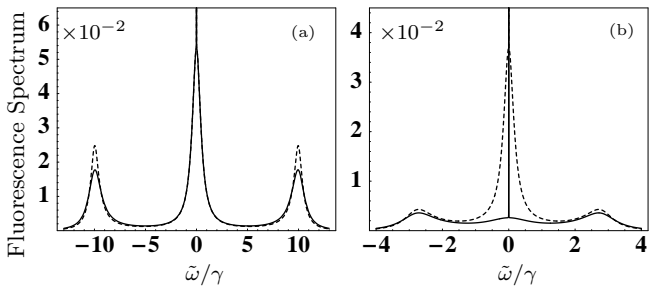}
\caption{\label{picture6} \small  Fluorescence spectrum for the degenerate system 
according to Eq.~(\ref{specpi}). 
The solid line (dashed line) shows
the spectrum with (without) the interference terms 
proportional to $\gamma_{12}$, $\gamma_{21}$. 
The Rayleigh peak (the vertical line at $\omega=\omega_L$) is present 
both with and without interference terms. 
Note that its weight is larger if the interference terms are 
taken into account. However, the sums of  the integrated coherent and incoherent spectra 
with and without the interference terms are identical, making the total intensity 
independent of the interference terms. 
In a), the parameters are 
$\Omega=5\times 10^7\,s^{-1}$, $\Delta=0$ and $\gamma=10^7\,s^{-1}$.
For b), we have  
$\Omega=10^7\,s^{-1}$, $\Delta=2\times10^7\,s^{-1}$ and $\gamma=10^7\,s^{-1}$. 
}
\ec
\end{figure}

In conclusion, the experimental observation of the fluorescence spectrum 
confirming the solid lines in Fig.~\ref{picture6} would give evidence for vacuum-mediated
interference effects as described by 
terms proportional to $\gamma_{12}$. 
So far, interference effects of this kind have not been observed in atomic systems.

\section{SPECTRUM OF RESONANCE FLUORESCENCE II - $ \sigma$ TRANSITIONS \label{sigma} }
This  Section is concerned with a brief discussion of the   fluorescence spectrum emitted on the $\sigma$ transitions. 
Since the laser field does not couple to these transitions, the spectrum contains only an incoherent part. 
We arrive at 
\be
S^{\sigma}(\tilde{\omega}) = \phi_{\sigma}\frac{b_{\sigma}\gamma}{\pi} \sum\limits_{i=3}^{4} 
\text{Re}\,\int\limits_{0}^{\infty} e^{-i\tilde{\omega}\tau}
\means{\delta \tilde{S}_i^+(\hat{t}+\tau) \delta \tilde{S}_i^-(\hat{t}) } \, d\tau  \;,
\label{s_sigma}
\ee
where $\phi_{\sigma}$ is a geometrical factor that we set equal to one in the following. It has been pointed out in 
Sec.~\ref{dscheme} that the light emitted on the $\sigma$ transitions  
is linearly polarized along $\mf{e}_x$ if the point of observation lies in  the  $y$-direction. Therefore, the cross terms 
\mbox{$\means{\delta \tilde{S}_3^+(\hat{t}+\tau) \delta \tilde{S}_4^-(\hat{t}) }$ } and 
\mbox{$\means{\delta \tilde{S}_4^+(\hat{t}+\tau) \delta \tilde{S}_3^-(\hat{t}) }$ } will, 
in principle, contribute to  the spectrum 
in Eq.~({\ref{s_sigma}). However, we find that the latter two-time averages are equal to zero. 
For different driving schemes where the laser field couples to the $\sigma$ transitions, 
the  cross-correlation terms  have to be taken into account 
as is the case in the work of  Polder et. al.~\cite{polder:76}.
The exact analytical expression for $S^{\sigma}$ is too bulky to  display it here.  Instead we will 
discuss $S^{\sigma}$ in the case of  the degenerate system ($B=\delta=0$) and for different regimes of 
the driving field strength that will be characterized by means of the saturation parameter 
\be
s = \frac{2|\Omega|^2}{\Delta^2 +  \gamma^2/4} \;.
\label{saturation}
\ee
In the range from a weak to a moderately strong laser field ($s<1$), 
a numerical analysis  reveals that $S^{\sigma}$  
can be written as 
\be
S^{\sigma}(\tilde{\omega})  \;\approx\;
     b_{\sigma}/b_{\pi}\,S_{\text{inc}}^{\pi} (\tilde{\omega}) + S_{\text{peak}}^{\sigma}(\tilde{\omega})   \;.
 \label{approx}
\ee
In this equation, the first term stands for the incoherent spectrum of a two-level atom according to Eq.~(\ref{two_level_spec}). 
The prefactor $b_{\sigma}/b_{\pi}$ accounts for the different branching probability of the $\sigma$ transitions as compared 
to the $\pi$ transitions.  The second term represents a narrow peak that is centered at the laser frequency $\omega=\omega_L$. It can 
be modeled as a Lorentzian of 
weight $\mc{W}_{\sigma}$ and width  $\Gamma_{\sigma}$,
\be  
S_{\text{peak}}^{\sigma}(\tilde{\omega})   = 
\frac{\mc{W}_{\sigma}}{\pi}\;\frac{\Gamma_{\sigma}}{\tilde{\omega}^2 + 
\Gamma_{\sigma}^2}\;.
\label{speak}
\ee
The weight of $S_{\text{peak}}^{\sigma}$ is determined by 
the total intensity emitted on the $\sigma$ transitions 
\be
I_{\text{st}}^{\sigma}=b_{\sigma} \gamma (\tilde{\vro}_{11} + \tilde{\vro}_{22})
\ee
and the weight of $b_{\sigma}/b_{\pi}\,S_{\text{inc}}^{\pi}$.  
We arrive at 
\be
\mc{W}_{\sigma} =4\, b_{\sigma} \gamma |\tilde{\vro}_{13}|^2 \,,
\ee
where $\tilde{\vro}_{13}$ is given in Eq.~(\ref{steady}). 
The width $\Gamma_{\sigma}$ of the additional peak is  smaller than the decay rate $\gamma$.  
If    $s$ is much smaller than unity, the width and the weight of $S_{\text{peak}}^{\sigma}$  
are given by 
\ba
\mc{W}_{\sigma}  &\approx & b_{\sigma}\frac{\gamma}{2} (1-2 s) s  \nonumber \\
 \Gamma_{\sigma} & \approx &  b_{\sigma}\frac{ \gamma}{4} \big[2- (2+b_{\sigma}) s\big]s  \;.
\ea
At the same time, the contribution of $S_{\text{inc}}^{\pi}\,b_{\sigma}/b_{\pi}$ 
to $S^{\sigma}$  is small  such that the    spectrum  
is dominated  by the central narrow peak $S_{\text{peak}}^{\sigma}$. 
If the field strength is increased, the weight of the extra peak $S_{\text{peak}}^{\sigma}$ gets 
smaller.   
Figure~\ref{picture7}~a) shows $S^{\sigma}$ (solid line) and  
\mbox{$S_{\text{inc}}^{\pi}\,b_{\sigma}/b_{\pi}$} (dotted line) for a moderately strong laser field, the saturation 
parameter is on the order of  unity. Nevertheless, the spectrum $S^{\sigma}$ is still dominated by the 
sharp peak $S_{\text{peak}}^{\sigma}$  that exceeds the central peak of the two-level spectrum by one order of magnitude. 
\begin{figure}[t!]
\bc
\includegraphics[scale=1]{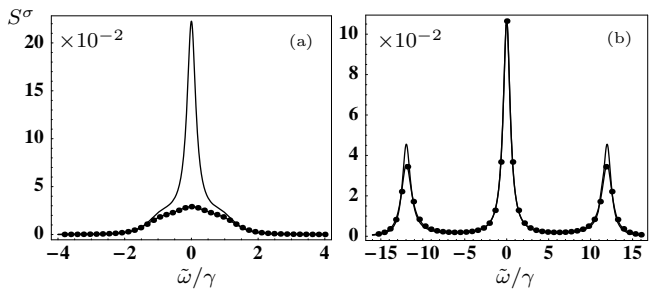}
\caption{\label{picture7} \small  Spectrum of 
resonance fluorescence emitted on the $\sigma$ transitions (solid line) 
in comparison with  the fluorescence spectrum of a  two-level atom (dotted line).  
The parameters in (a) are \mbox{$\Omega = 5\times 10^6\,s^{-1}$},   
\mbox{$\Delta =6\times 10^6\,s^{-1}$} and \mbox{$ \gamma = 10^7\,s^{-1}$}. 
 In (b), the parameters are 
$\Omega = 6\times 10^7\,s^{-1}$,   $\Delta = 0$ and $ \gamma = 10^7\,s^{-1}$.  
Note that  $S^{\sigma}$ deviates slightly from the  
Mollow triplet in the region of the sideband peaks.}
\ec
\end{figure}

For a strong driving field ($s \gg 1$), the weight of $S_{\text{peak}}^{\sigma}$ goes to zero 
and the central peak  of  $S^{\sigma}$ coincides with the corresponding peak of the Mollow spectrum. 
However, the  sideband peaks of $S^{\sigma}$ differ from 
those of a two-level atom as can be seen from Fig.~\ref{picture7}~b). 
In the secular limit, it is advantageous to employ the dressed state picture in order 
to obtain analytic expressions for the sideband peaks, being  well separated from the central peak 
whose  analytic form can be taken over from the well-known results for a two-level atom~\cite{kimble:76, tannoudji:api}.   
The fluorescence spectrum for a resonant driving field 
can be achieved by a tedious but straightforward calculation  that follows  the 
procedure of Chapter \mbox{VI.E} in~\cite{tannoudji:api}, 
\ba
S^{\sigma}(\tilde{\omega}) & \approx & 
\gamma\frac{b_{\sigma}}{8\pi} \; \frac{\Gamma_{\text{sb}}}
{\Gamma_{\text{sb}}^2 +(\Omega_1 - \tilde{\omega})^2} 
 \label{resonant_sigma} \\[0.2cm]
&& + \,\gamma\frac{b_{\sigma}}{4\pi}\;\frac{\gamma/2}{\gamma^2/4 +\tilde{\omega}^2} 
 +\gamma \frac{b_{\sigma}}{8\pi}  \;\frac{\Gamma_{\text{sb}}}
{\Gamma_{\text{sb}}^2 +(\Omega_1 + \tilde{\omega})^2}\nonumber \;,
\ea
where $\Omega_1=\sqrt{4|\Omega|^2 +\Delta^2}$.
A comparison of the latter equation with the corresponding expression for the Mollow spectrum reveals that   
the weights of the sideband peaks differ only by the branching probability $b_{\sigma}$. 
For the width of the sideband peaks in Eq.~(\ref{resonant_sigma}) we find 
\be
\Gamma_{\text{sb}} = \frac{1}{4}\sqrt{\gamma_1\gamma_2}  +\frac{\gamma}{2}
=\frac{1}{4}(3-b_{\sigma})\gamma\,.
\label{sideband_width}
\ee
Note that the second equality  is obtained by virtue of Eq.~(\ref{branching_prob}) .
The ratio between the heights of the central peak at $\tilde{\omega}=0$ and the sideband peaks at 
$\tilde{\omega}=\pm \Omega_1$ is found to be   $3-b_{\sigma}$. For $b_{\sigma}=2/3$, the peak ratio is thus  
$3:7:3$. By contrast, the peak ratio of the Mollow spectrum reads $1:3:1$.  
A precise measurement of the peak ratio  would thus provide a means of determining the branching probability $b_{\sigma}$ 
of the degenerate system experimentally. 

Note that the width of the sideband peaks in Eq.~(\ref{sideband_width}) 
depends on the cross-damping terms  $\sqrt{\gamma_1\gamma_2}$  that 
appear in the master equation through the spontaneous emission term $\mc{L}_{\gamma}\tilde{\vro}$ in Eq.~(\ref{liou}). 
If these interference terms were not present, the peak ratio would not depend on the 
branching probabilities and would be given by $1:2:1$.  The spectrum emitted on the $\sigma$ transitions shows thus an 
indirect signature of interference. 

\section{Discussion   \label{discussion}}
In   Section~\ref{sec_pi} we have shown that  the interference terms proportional 
to $\gamma_{12}$ contribute only to the spectrum of resonance fluorescence, 
but not to the  total intensity in Eq.~(\ref{intensity2}). 
Here we demonstrate  that  
this result is a consequence of the principle of complementarity, applied to time and energy. 

If  the total intensity is measured, 
complementarity does not impose any restrictions on the time resolution of 
the measurement since the photon energies are not observed. 
It is thus possible to observe the temporal aspect of the radiative 
cascade, i.e. one could determine the photon emission times. 
The time evolution of the driven atom is then most suitably described in the bare state basis. 
For example, assume that  the atom is initially in ground state $\ket{3}$.  The laser field 
will induce Rabi oscillations between the excited state $\ket{1}$ and $\ket{3}$.  
Immediately after the  spontaneous emission of 
a photon, the atom is found in ground state $\ket{3}$ ($\pi$ transition) or $\ket{4}$ 
($\sigma$ transition). Subsequently, this sequence of Rabi oscillations and a spontaneous 
emission event is repeated. 
In this description, each emission process on one of the  $\pi$ transitions is independent of the 
other $\pi$ transition. 
However, quantum interference does only occur if various 
indistinguishable transition amplitudes connect a common initial state  to a common final state. 
Since the $\pi$ transitions do neither share a common initial nor a 
common final state,  we must conjecture that the total intensity is not affected by interference. 

The lack of interference in the total intensity can also be explained by 
drawing  an analogy  to the two-slit experiment.  
It is well known that the interference pattern vanishes as soon as 
it is principally possible to know through which of the two slits 
each object (electrons or photons) has moved. 
Similarly, the internal states of our atom can be regarded as a  which-way marker. 
Since the experimental conditions allow, at least in principle, to
determine the  atomic ground state immediately after the detection of a $\pi$-photon, 
one could decide on which of the two $\pi$-transitions the photon was emitted. 
Consequently, the observer could  reveal the 
quantum path taken by the system 
and hence there  is no signature of interference. 
Note that this argument requires that the retardation between the 
times of emission and detection is much smaller 
than the time between successive emissions.  
This condition can typically be achieved in atomic systems.


A totally different situation arises if the detector measures the 
spectrum of resonance fluorescence and hence the energy of the emitted 
photons.  First of all, it is advantageous to 
illustrate the energy aspect of the   cascade of spontaneously emitted photons 
in the dressed state picture
(see Fig.~\ref{picture8})  rather than in the 
bare state picture. 
The crucial difference between the measurement of the total 
intensity and the fluorescence spectrum is the following. 
In the latter case, the observer decides to determine  
the photon energies precisely. Since time and energy are 
complementary observables, the temporal aspect of 
the radiative cascade is not  accessible simultaneously. 
Next we demonstrate that precisely this lack of information about the 
temporal sequence of photon emissions allows for 
the interference mechanism in the fluorescence spectrum. 

A quantitative description of time-energy complementarity is achieved 
via the time-energy uncertainty. 
If the photon energies are determined with a precision of $\Delta\omega$, the  
time-energy uncertainty relation enforces that the time of observation has to be at least on the order of $1/\Delta\omega$. 
Since the observer can only notice the detection of a photon  after the observation time has elapsed, 
the photon emission times are indeterminate within a time interval 
of \mbox{$\Delta t=1/\Delta\omega$}. 
For the moment we envisage  an ideal measurement of 
the fluorescence spectrum. In this case, the atom will emit (infinitely) many photons 
during the (infinite) time of observation. 
In addition, the photon emission times are  indeterminate, 
and thus the time order in which these photons have been emitted  is unknown.  
It follows that the transition amplitudes corresponding to the various time orderings of the photons 
will interfere. 

We  illustrate this interference mechanism on the basis of Fig.~\ref{picture8} 
that shows   a cascade of only  two photons, 
one emitted on a $\pi$ transition and the other on a $\sigma$ transition. 
Assume that the atom is initially in  the dressed state 
$\ket{4(N)}$.  In one of the two cascades, the atom decays first to 
the state $\ket{4(N-1)}$  by the emission of a $\pi$ photon on the 
bare state transition $\ket{2}\rightarrow\ket{4}$.  The subsequent emission of 
a $\sigma$ photon takes the atom to the state $\ket{1(N-2)}$ within the 
manifold with $N-2$ laser photons. In the second cascade, the time order of the 
two photons is reversed. The atom decays now first to 
the state $\ket{1(N-1)}$  by the emission of a $\sigma$ photon, and 
then to the final state $\ket{1(N-2)}$ under the emission of a $\pi$ photon. 
In contrast to the first cascade, this $\pi$ photon is now 
emitted on the bare state transition $\ket{1}\rightarrow\ket{3}$.  
Since the two cascades in Fig.~\ref{picture8} have the same initial and final states, 
and since it is in principle impossible to determine the quantum path taken by the system, the two transition amplitudes 
corresponding to different time orders of photon emissions interfere. 
In one of the transition amplitudes  the $\pi$ photon  stems from the 
$\ket{2}\rightarrow\ket{4}$ transition, and in the other from the  $\ket{1}\rightarrow\ket{3}$ transition.  
Exactly this mechanism  gives rise to the interference effects in the fluorescence spectrum that are mediated 
by the cross-damping terms in Eq.~(\ref{specpi}). 
\begin{figure}[t!]
\includegraphics[scale=1]{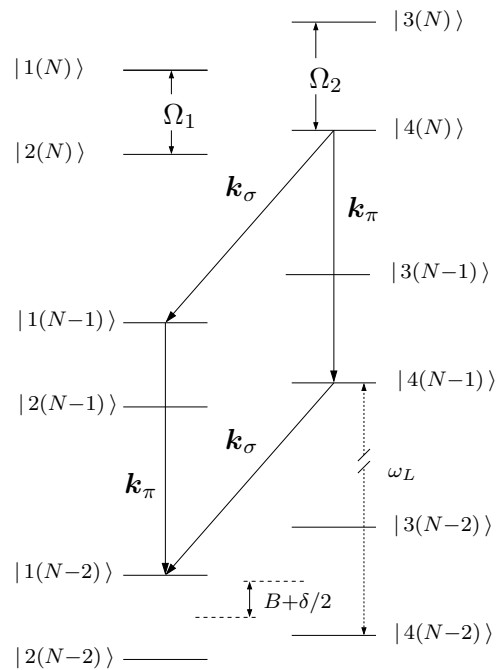}
\caption{\label{picture8} 
Radiative cascade in the dressed states of the system [see Eqs.~(\ref{dressed1}) and (\ref{dressed2})]. 
The splitting of the dressed states for a fixed number of laser photons $N$ is not to scale. 
Each of the two indicated cascades involves the emission of 
 a $\pi$ photon  and a $\sigma$ photon  with wave vector 
$\mf{k}_{\pi}$ and $\mf{k}_{\sigma}$, respectively. 
Depending on the time order of  their 
emission, the $\pi$ photon  is either emitted on transition 
$\ket{4(N)}\rightarrow \ket{4(N-1)}$ 
or $\ket{1(N-1)}\rightarrow \ket{1(N-2)}$, corresponding to the bare state transitions
$\ket{2}\rightarrow\ket{4}$ and $\ket{1}\rightarrow\ket{3}$, respectively.
Since the final and initial states of the two cascades are identical, 
the corresponding transition amplitudes may interfere. 
}
\end{figure}

The provided explanation can also be employed to illustrate why there is no interference in the 
fictitious situation of perpendicular dipole moments $\mf{d}_1$ and $\mf{d}_2$. In this 
case, a photon can either stem from $\mf{d}_1$ or $\mf{d}_2$, but not from both transitions.  
It is then impossible to realize both cascades in  Fig.~\ref{picture8}, and 
hence there is no interference.  
Moreover, it becomes now clear why the spectrum emitted on the $\sigma$ transitions depends on 
the interference terms $\gamma_{12}$ and $\gamma_{21}$. For anti-parallel dipole moments $\mf{d}_1$ and $\mf{d}_2$ 
there are two transition amplitudes that involve the emission of a $\sigma$ photon , and for perpendicular dipole moments  
there would be only one.  We emphasize that the discussion has been restricted to  a cascade of only two photons  for 
the sake of simplicity. 
In principle, all possible cascades with an arbitrary number of photons have to be considered, but the 
general idea remains the same. 

It is also possible to provide an explanation for the interference in the coherent  spectrum, 
but the elastic scattering events  cannot be visualized in the dressed state 
basis. However, in  the case of low saturation ($s\ll1$) 
the process of elastic scattering  can  be illustrated in the bare state basis such that the atom 
hops from one ground state to another  by the absorption of a laser photon and the  emission of a scattered photon. 
The excited states act as intermediate states and can be adiabatically eliminated. 
Since it is impossible to tell on which of the two  $\pi$ transitions the photon  was scattered, 
it is plausible that one has to sum the  scattering amplitudes 
first and then take the absolute value squared in order to obtain the  weight of the Rayleigh line  in Eq.~(\ref{cohline}).  


Next we demonstrate how  the interfering transition amplitudes  that correspond to different   time orders of photon emissions  
enter the expression for the spectrum of resonance fluorescence in Eq.~(\ref{specpi}).  
Let $a_{\pi}$ ($a_{\pi}^{\dagger}$) be the annihilation (creation) operator of a photon in 
a mode of the radiation field that is actually observed by the detector, being  
sensitive only to photons 
emitted on the $\pi$ transitions. The rate at which the photon number in 
this particular mode changes is  given by 
\be
R_{\pi}(t) =\partial_t\,\mean{a_{\pi}^{\dagger}(t)a_{\pi}(t)}\;.
\label{interpretation}
\ee
If one follows the lines of Chapter 7 in~\cite{agarwal:qst}, one can show that the steady-state value of 
$R_{\pi}$ is proportional to the spectrum of resonance fluorescence,  
\be
\lim\limits_{t\rightarrow\infty}R_{\pi}(t)  \sim S^{\pi}(c\,|\mf{k}_{\pi}|-\omega_L)\;. 
\label{int2}
\ee
In this equation, $\mf{k}_{\pi}$ denotes the  wave vector that corresponds to the observed 
mode $a_{\pi}$, and $c$ is the speed of light. 
In order to evaluate the left hand side of Eq.~(\ref{int2}), 
we will label the basis states 
\mbox{$\ket{i(N);\,{\{n\}}}$} of the total system (atom + laser field + vacuum modes) 
by three quantum numbers, namely the dressed states $i$, the number of laser photons $N$ and 
the state of the vacuum modes $\{n\}$.  
The mean value on the right hand side of Eq.~(\ref{interpretation}) becomes then 
\be
\mean{a_{\pi}^{\dagger}(t)a_{\pi}(t)} =
\sum\limits_{i=1}^4\sum\limits_{N,\,\{n\}} |C_{N,\{n\}}^{i}(t)|^2\,N_{\pi}(\{n\})\;,
\ee
where $|C_{N,\{n\}}^{i}(t)|^2$ is the probability to find the system at time $t$ in state 
\mbox{$\ket{i(N);\,{\{n\}}}$} and $N_{\pi}(\{n\})$ is the expectation value of 
 $a_{\pi}^{\dagger}a_{\pi}$ in this
state. We assume that the system is  in some initial state $\ket{\psi_0}$ at time $t=0$ with 
all vacuum modes being empty. 
If the time evolution operator is labeled by $U(t,0)$, the transition amplitude from the initial state 
$\ket{\psi_0}$ to the final state \mbox{$\ket{i(N);\,{\{n\}}}$} can be written as 
\be
C_{N,\{n\}}^{i}(t) = \bra{i(N);\,{\{n\}}}U(t,0)\ket{\psi_0}\;. 
\label{amplitude}
\ee
Let us assume that the final state contains  $q$ scattered photons that are characterized 
by their wave  and polarization vectors,
\mbox{$\{n\}=\{\mf{k}_{\pi}\mf{\eps}_{\pi},\,\mf{k}_2\mf{\eps}_2,\ldots,\,\mf{k}_q\mf{\eps}_q\}$}.  
We do not attempt to evaluate Eq.~(\ref{amplitude}) explicitly, but in principle one 
would introduce $q-1$ intermediate 
states and arrange the $q$ scattered photons into a certain order. But since there is no 
distinguished time order of the scattered photons,  there 
are, in principle, $q!$ transition amplitudes involved in the evaluation of Eq.~(\ref{amplitude}) 
that will all interfere.  

 In conclusion, we demonstrated that the interference in the spectrum from the $\pi$ transitions 
can be explained in terms of interference between transition amplitudes that correspond to 
different time orders of photon emissions. If the spectrum of resonance fluorescence 
is observed, the principle of complementarity enforces that these transition amplitudes are  indistinguishable. 
If the total intensity is recorded by a broadband detector, the temporal aspect of the radiative cascade 
can in principle be observed. Consequently, the possibility of interference between different time orders of photon emissions  
is  ruled out. 
The preceding discussion of our results also implies that  the experimental setup---potentially after the
photon emissions---decides if interference takes place,  
a feature that is also known from quantum eraser schemes~\cite{kim:00,walborn:02}. 

We now  refine our analysis and consider a detector with a finite frequency resolution $\Delta\omega$ that   
allows us to study the continuous transition from perfect frequency resolution to perfect time resolution. 
For simplicity, we consider only the degenerate system ($B=\delta=0$). 
If a filter of bandwidth $\lambda$ 
and setting frequency $\omega$ is placed in front of 
a broadband detector, 
the spectrum can be determined with an accuracy of $\lambda$, and the 
temporal resolution is on the order of $\lambda^{-1}$. 
The spectrum of resonance fluorescence  emitted on the $\pi$ transitions reads then~\cite{eberly:77} 
\ba
&& \hspace*{-0.8cm} S^{\pi}(\tilde{\omega},\lambda) =   \label{specpifilter} \\[0.2cm] 
&& \hspace*{-0.8cm} \frac{1}{\pi}\, \sum\limits_{i,\,j=1}^{2} \gamma_{ij} 
\,\text{Re} \int\limits_{0}^{\infty} e^{-i\tilde{\omega}\tau} e^{-\lambda\tau}
\means{\tilde{S}_i^+(\hat{t}+\tau)\tilde{S}_j^-(\hat{t})}\, d\tau  \;.\nonumber
\ea
In the absence of interference terms the spectrum   will be denoted by $S_0^{\pi}(\tilde{\omega},\lambda)$ 
and is obtained from Eq.~(\ref{specpifilter}) by omitting the  terms proportional to $\gamma_{12}$ and $\gamma_{21}$.  
For the rest of this Section we assume that the saturation parameter $s$ is much smaller than unity. 
To a first approximation, the incoherent contribution to the 
spectrum with interference terms can then be neglected. In the presence of the 
filter, the coherent $\delta$-peak becomes a   Lorentzian of width $\lambda$ 
and weight $I_{\text{coh}}^0 + I_{\text{coh}}^{\text{int}}$,  and thus 
we obtain 
\be
S^{\pi}(\tilde{\omega},\lambda) \approx 
\frac{I_{\text{coh}}^0 + I_{\text{coh}}^{\text{int}}}{\pi}\,
\frac{\lambda}{\tilde{\omega}^2 + \lambda^2} \,. \label{spilambda}
\ee
Similarly, we neglect the  contribution of $S_{\text{inc}}^{\pi}$  to the 
spectrum without interference terms in Eq.~(\ref{specpiperpapprox}), 
the $\delta$-peak 
becomes  a Lorentzian of width $\lambda$ and weight $ I_{\text{coh}}^0$, 
and   $S_{\text{peak}}^{\pi}$  is replaced by a Lorentzian of 
width  $\Gamma_{\pi}+\lambda$ and weight $I_{\text{coh}}^{\text{int}}$, 
\be
S_0^{\pi}(\tilde{\omega},\lambda)   \approx   
\frac{I_{\text{coh}}^0 }{\pi}\,  \frac{\lambda}{
\tilde{\omega}^2 + \lambda^2}   
  \,+\, \frac{I_{\text{coh}}^{\text{int}}}{\pi}\,
\frac{ \Gamma_{\pi} +\lambda}{\tilde{\omega}^2 + (\Gamma_{\pi}+\lambda)^2}   \;. \label{spilambdaperp}
\ee
Figure~\ref{picture9} shows the fluorescence spectrum according to Eq.~(\ref{specpifilter}) (solid lines) for 
different values of the filter bandwidth $\lambda$ and for low saturation.  The dashed lines are the spectra without 
the interference terms. In Fig.~\ref{picture9}~a), the bandwidth $\lambda$  is much smaller than  $\Gamma_{\pi}$. 
Therefore, the widths of the  lines  $S^{\pi}(\tilde{\omega},\lambda)$ and $S_0^{\pi}(\tilde{\omega},\lambda)$ 
are clearly distinct. If $\lambda$ is increased, the differences between the spectra with and 
without the interference terms diminish until both curves are virtually identical for $\lambda=\gamma$ 
(Fig.~\ref{picture9}~d)).  
\begin{figure}[t!]
\bc
\includegraphics[scale=1]{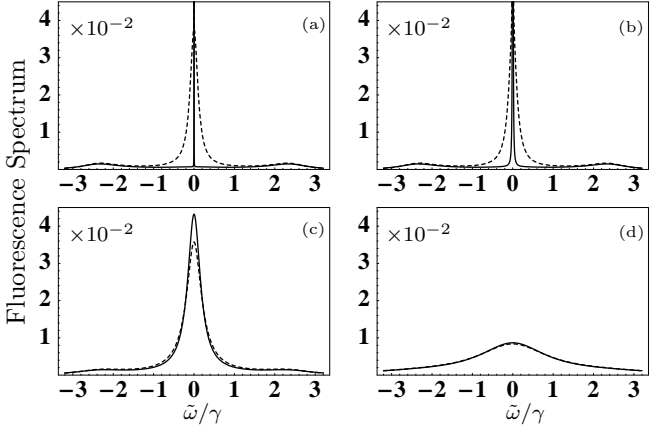}
\caption{\label{picture9} \small     
The solid lines show the fluorescence spectra recorded with a finite frequency resolution $\lambda$. The 
dashed curves are the spectra without the interference terms proportional to $\gamma_{12}$, $\gamma_{21}$ 
in Eq.~(\ref{specpifilter}). The parameters are 
$\Omega = 7\times 10^6 s^{-1}$,  $\Delta=2\times 10^7 s^{-1}$, $\gamma = 10^7 s^{-1}$ 
and $B=\delta=0$.  
This corresponds to a  saturation parameter of \mbox{$s=0.235$} and a   
mean number of photons per unit time of approximately $9.4\times 10^5 s^{-1}$. 
The filter bandwidths are given by a)~$\lambda=10^2 s^{-1}$, 
b)~$\lambda=10^4 s^{-1}$,
c)~$\lambda=1.9\times 10^6 s^{-1}$ and 
d)~$\lambda=10^7 s^{-1}$.  
}
\ec
\end{figure}

These results can be understood as follows. 
With an increasing filter bandwidth $\lambda$, the smallest time interval $\Delta t$ 
that can be resolved by the detector without violating the time-energy uncertainty gets shorter. 
Therefore, the observer can in principle obtain more information about 
the quantum path taken by the atom.  
Consequently, we  expect that the signature of interference in the fluorescence spectrum 
diminishes for increasing $\lambda$. This is in agreement with Fig.~\ref{picture9}  
and completely analogous to a two-slit experiment, where the visibility of the interference pattern is reduced at the 
cost of which-path information and vice versa~\cite{englert:96}.

Furthermore, we demonstrate that the time-energy uncertainty relation allows to estimate the smallest filter 
bandwidth  $\lambda$ 
for which the spectra with and without interference terms should be indistinguishable. 
Since the total number of photons emitted per unit time is equal to 
$\gamma(\tilde{\vro}_{11} + \tilde{\vro}_{22})$, the mean time 
between successive photon emissions is determined by $\bar{\Theta}=1/[\gamma(\tilde{\vro}_{11} + \tilde{\vro}_{22})]$.   
If the  bandwidth $\lambda$ is chosen such that the temporal resolution could be much better than the mean time between 
successive photon emissions, i.e. 
\mbox{$\lambda^{-1}  \ll  \bar{\Theta}=1/[\gamma(\tilde{\vro}_{11} + \tilde{\vro}_{22})]$}, 
we have 
\be
\lambda \gg \gamma (\tilde{\vro}_{11} + \tilde{\vro}_{22})\approx (1-s) s \gamma/2\;. \label{ineq3}
\ee
Under these conditions, the radiative cascade of photons could be observed in a time resolved way and 
it is extremely unlikely that more than one spontaneous emission  takes place during the time of observation. 
Since this rules out the interference mechanism as 
described in Sec.~\ref{discussion}, 
the signature of interference in the fluorescence spectrum should disappear. 
But if inequality~(\ref{ineq3}) holds, it follows that 
$\lambda \gg \Gamma_{\pi}$, 
and in this case  $S^{\pi}(\tilde{\omega},\lambda)$ and 
$S_0^{\pi}(\tilde{\omega},\lambda)$ are indeed   
indistinguishable as can be seen from Eqs.~(\ref{spilambda}) and (\ref{spilambdaperp}).  
This is confirmed by Fig.~\ref{picture9}~d)   
that shows  $S_0^{\pi}(\tilde{\omega},\lambda)$  and $S^{\pi}(\tilde{\omega},\lambda)$ 
for a bandwidth $\lambda$ that is about ten times larger than  
the mean number of photons emitted per unit time. 
The two spectra are now virtually indistinguishable. 

\begin{figure}[b!]
\bc
\includegraphics[scale=1]{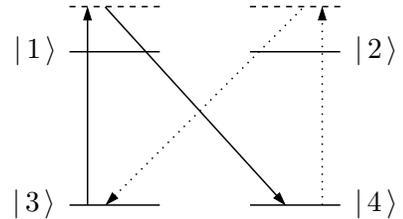}
\caption{\label{picture10} \small   Schematic representation of elastic scattering events into the 
\mbox{$3\rightarrow1\rightarrow4$} (solid arrows) 
and   $4\rightarrow2\rightarrow3$ channels (dotted arrows). 
These processes 
account for the sharp peak in the fluorescence spectrum $S^{\sigma}$ emitted on the $\sigma$ transitions. 
  }
\ec
\end{figure}
It remains to explain the sharp peaks in the incoherent spectrum. To this end we return to Fig.~\ref{picture4} 
that shows the incoherent spectrum $S_{\text{inc}}^{\pi}$ for several values of the parameter $\delta$. 
A narrow central peak occurs only in case of the non-degenerate system ($\delta\not= 0$), and thus only if 
the weight of the Rayleigh line deviates from its maximal value attained at $\delta=0$. 
Therefore, the narrow central peak in the incoherent spectrum may  be regarded  as a 
(partially)  broadened coherent peak.  This broadening can be understood as follows. Except for $\delta=0$, the 
two $\pi$ transitions are not equivalent since the absolute value and the phase of  the coherences 
$\means{\tilde{S}_1^+}$ and $\means{\tilde{S}_2^+}$ will be different. 
The time that the atom spends on the $1\leftrightarrow3$ transition can thus be regarded as a dark period with 
respect to the  $2\leftrightarrow4$ transition and vice versa. This suggests that the 
sharp peaks in the incoherent spectrum can be explained in terms of electron shelving~\cite{hegerfeldt:95,plenio:98,evers:02}. 
This explanation is also applicable to the sharp peak in the spectrum from the $\sigma$ transitions. 
Figure~\ref{picture10} illustrates the scattering events that give rise to 
this peak. If the  atom is initially in state $\ket{3}$, a scattering event can bring it to 
state $\ket{4}$ (solid arrows). The scattered photon has then been emitted on one of the $\sigma$ transitions. 
Before the next photon can be scattered on that same transition, a similar scattering process has to take place 
into the $4\rightarrow2\rightarrow3$ channel (dotted arrows).  Consequently, every emission on one  of the  
$\sigma$ transitions   is followed by a dark period on that same transition.

It should be mentioned  that  related interference effects between transition amplitudes corresponding to 
different time orders of photon emissions do also play a role 
in the fluorescence spectrum of other systems~\cite{oscillator}. 
However, the distinguished feature of the system presented here is that 
this mechanism gives rise to interference effects between the two $\pi$ transitions 
that do not share a common state. 

We also  point out that our system belongs to a class of setups that display interference and complementarity 
in the time-energy domain. In a conventional double-slit experiment, spatially 
separated pathways result in an interference pattern in position space.  
This is in contrast to our setup, where  different temporal paths lead to interference in the energy domain.  
The work presented here is thus related to recent   double-slit experiments in the time-energy 
domain~\cite{lindner:05, wollenhaupt:05}. 
In these experiments, ultra-short laser pulses of 
atto- or femtosecond duration open different time windows for the photoionization of an atom. 
If the energy spectrum of the photoelectrons is measured, these time-slits are indistinguishable and 
an interference pattern is observed. 
Moreover, it has been demonstrated that interference in the time-energy domain can occur 
in the intensity correlations of different spectral components in a two-level atom~\cite{aspect:80, schrama:91, schrama:92}.

\section{SUMMARY \label{sum}}
The key result of this paper is that there is quantum interference in the 
spectrum of resonance fluorescence  under conditions of no interference in 
the total intensity, being enforced by the principle of complementarity. For 
the system considered here, it claims that it is impossible to observe the 
temporal and the energy aspect of the radiative cascade of the 
atom at the same time.  If the fluorescence spectrum is observed, 
the photon emission times are indeterminate. The interference in the 
fluorescence spectrum can thus be explained in terms of  interferences 
between transition amplitudes that correspond  to different time orders 
of photon emissions. 

It has been shown that the degree of interference in the fluorescence 
spectrum emitted on the $\pi$ transitions can be controlled by means of an 
external  magnetic field. In particular, the degree of interference in the coherent 
part of the spectrum can be adjusted from perfect constructive to perfect 
destructive interference. Under conditions of perfect destructive interference, 
the weight of the Rayleigh line is  completely suppressed.  
If the difference $\delta$ between the resonance frequencies of the 
$\pi$ transitions is different from zero, the incoherent 
spectrum emitted on the $\pi$ transitions contains a 
very narrow peak whose width is smaller than the 
decay rate $\gamma$. This peak has been identified 
as a partially broadened coherent peak and can be 
explained in terms of electron shelving. 

The spectrum emitted on  the $\sigma$ transitions contains only an 
incoherent part. In the case of a weak driving field and for the degenerate 
system, the fluorescence spectrum displays a narrow peak that can be 
regarded as  broadened coherent  peak.  
For a strong driving field, the widths of the  sideband peaks 
differ from the Mollow spectrum. We have shown that  
the ratio between the peak heights of the central and the sideband 
peaks display an indirect signature of interference. 
In addition, a   measurement of the  relative  peak heights  allows to 
determine the branching probability $b_{\sigma}$  
of the spontaneous decay of each excited 
state into the  $\sigma$ channel. 

\begin{acknowledgments}
MK thanks  Z. Ficek for helpful discussions. 
\end{acknowledgments}

\appendix*

\section{CALCULATION OF THE TWO-TIME AVERAGES \label{sectwotime}}
In this section we outline how  the functions
\be
S_{ij}(\tilde{\omega}) =  \text{Re}\,\int\limits_{0}^{\infty} e^{-i\tilde{\omega}\tau}
\means{\delta \tilde{S}_i^+(\hat{t}+\tau) \delta \tilde{S}_j^-(\hat{t}) } \, d\tau  \;.
\label{twotime}
\ee
can be evaluated by means of the quantum regression theorem~\cite{lax:63,carmichael:bk}.  
To this end we introduce the operators $\mc{A}_{ij}$ that 
are connected to  the atomic transition  operators $A_{ij}$
(taken in the Schr\"odinger picture) by
\be
\mc{A}_{ij}= W^{\dagger} A_{ij} W  \;, 
\label{mca}
\ee
where the unitary transformation $W$ is defined in Eq.~(\ref{trafo_w}). 
In particular, the operators $\tilde{S}_i^{\pm}$ introduced in Sec.~(\ref{dscheme}) can be 
identified with the operators $\mc{A}_{ij}$ according to 
\be
\tilde{S}_1^+ = \mc{A}_{13}\quad \tilde{S}_2^+ = \mc{A}_{24}
\quad \tilde{S}_3^+ = \mc{A}_{23} \quad \tilde{S}_4^+ = \mc{A}_{14} \;.
\ee
The corresponding  Heisenberg operators  are then defined as 
\be
\mc{A}_{ij}(t) = U^{\dagger}(t,0)\, \mc{A}_{ij}\,U(t,0)\;,
\ee
and the time evolution operator has been labeled by $U$. 
A straightforward calculation shows that the mean values of the these Heisenberg operators 
are   directly related to   the matrix elements 
of the reduced density operator $\tilde{\vro}$ in the rotating frame, 
\be
\mean{\mc{A}_{ij}(t)}  = 
 \text{Tr}_{\text{A}} \big[   A_{ij}  \tilde{ \vro} (t)  \big] = \tilde{\vro}_{ji}(t) \;.
\ee
In this equation,   $\text{Tr}_{\text{A}}[\,\cdot\,]$ denotes 
the trace over  atomic degrees of freedom.
Next we arrange the operators $\mc{A}_{ij}$ in a column vector  
\ba
\mf{L} & = & ( \mc{A}_{11}, \mc{A}_{21}, \mc{A}_{31}, \mc{A}_{41}, \mc{A}_{12}, 
\mc{A}_{22}, \mc{A}_{32}, \mc{A}_{42}, \nonumber\\[0.2cm]
&&\hspace*{2cm}\mc{A}_{13}, \mc{A}_{23}, \mc{A}_{33}, \mc{A}_{43}, \mc{A}_{14}, \mc{A}_{24}, \mc{A}_{34})^t
\nonumber
\ea
such  that $\mean{\mf{L}(t)}$ coincides with the Bloch vector  $\mf{R}(t)$ of 
Eq.~(\ref{bloch_vector}), i.e. $\mean{\mf{L}(t)} = \mf{R}(t)$.  
It follows that the mean values $\mean{\mf{L}(t)}$ obey the generalized Bloch Equation~(\ref{bloch_eq}).  
If we decompose each component of $\mf{L}$ in mean values an fluctuations according to 
$\mc{A}_{ij} = \delta \mc{A}_{ij} +  \means{\mc{A}_{ij}}\hat{1}$, we can cast 
$\mean{\mf{L}}$ into the form 
\be
\mean{\mf{L}(t)} = \mean{\delta \mf{L }(t)} + \means{\mf{L} }\;, 
\label{dl}
\ee 
where $\means{\mf{L}} = \mf{R}_{\text{st}} = -\mc{M}^{-1}\mf{I}$. If 
Eq.~(\ref{dl}) is plugged into Eq.~(\ref{bloch_eq}) we obtain a 
homogeneous equation of motion for the fluctuations, 
\be
\partial_{t}\mean{\delta \mf{L}(t)}= \mc{M}\,\mean{\delta \mf{L}(t)} \;.
\ee
The two-time correlation functions $\mean{\delta  L_i(\hat{t} + \tau)\delta L_j(\hat{t})}$  for 
$i \in \{1,\ldots,15\}$ and fixed $j$ can be written in vector notation as 
\mbox{$\mf{g}^{j}(\hat{t},\tau) = \mean{\delta \mf{L}(\hat{t} + \tau)\delta L_j(\hat{t})}$}. 
According to the quantum regression theorem, 
$\mf{g}^{j}$ obeys the same equation of motion than 
the one-time averages $\mean{\delta \mf{L}(t)}$, 
\be
\partial_{\tau} \mf{g}^j = \mc{M}\,\mf{g}^j  \qquad \text{for}\quad\tau\ge 0\;.
\ee
If $\mf{G}^j(\hat{t},z)$ denotes the   Laplace transform of $\mf{g}^{j}(\hat{t},\tau)$ with 
respect  to $\tau$, it follows  
\be
\mf{G}^{j}(\hat{t},z) = \left[  z\, \hat{1} - \mc{M} \right]^{-1} \mf{g}^{j}(\hat{t},0) \;.
\ee
We need the Laplace transform at $z=i\tilde{\omega}$ in steady state to 
determine the functions $S_{ij}(\tilde{\omega})$ of Eq.~(\ref{twotime}). 
With the definitions 
\be
 \mf{R}^j = \lim\limits_{\hat{t}\rightarrow \infty} \mf{g}^{j}(\hat{t},0)   
\quad \text{and}\quad 
\mf{K}^j(\tilde{\omega}) = 
\lim\limits_{\hat{t}\rightarrow \infty} \mf{G}^{j}(\hat{t},z=i\tilde{\omega})
\ee
we arrive at
\be
\mf{K}^j(\tilde{\omega}) = \big[ i\tilde{\omega}\, \hat{1} - \mc{M} \big]^{-1}\mf{R}^j \;.
\label{final}
\ee
The relevant  correlation functions that are needed for the evaluation of Eq.~(\ref{s_pi}) and (\ref{s_sigma}) are then 
given by 
\be
\begin{array}{l@{\,=\,}l@{\hspace*{0.4cm}}l@{\,=\,}l}
 S_{11}(\tilde{\omega}) & \text{Re}\left[\mf{K}^3(\tilde{\omega})\right]_{9}  &
S_{21}(\tilde{\omega})&  \text{Re}\left[\mf{K}^3(\tilde{ \omega})\right]_{14} \\[0.3cm]
 S_{22}(\tilde{\omega}) &  \text{Re}\left[\mf{K}^8(\tilde{\omega}) \right]_{14} &
 S_{12} (\tilde{\omega})&  \text{Re}\left[\mf{K}^8 (\tilde{\omega})\right]_{9} \\[0.3cm]
  S_{33}(\tilde{\omega}) &  \text{Re}\left[\mf{K}^7 (\tilde{\omega})\right]_{10} &
   S_{44}(\tilde{\omega}) &  \text{Re}\left[\mf{K}^4(\tilde{\omega})\right]_{13}  \;.
 \end{array}
 \label{correspond} 
 \ee
Finally, we remark that Eq.~(\ref{specpifilter}) can be evaluated if 
one replaces $i\tilde{\omega}$  in Eq.~(\ref{final}) by $i\tilde{\omega}+\lambda$.

\end{document}